\documentclass{aa}
\usepackage{epsf}
\usepackage{graphicx}
\begin{document}
   \title{Evolution of V838 Monocerotis \\
during and after the 2002 eruption}

   \author{R. Tylenda}

   \offprints{R. Tylenda}

   \institute{Department for Astrophysics, 
              N. Copernicus Astronomical Centre,
              Rabia\'nska 8, 87-100 Toru\'n, Poland\\
              \email{tylenda@ncac.torun.pl}}

   \date{Received   }

   \abstract{By fitting the available photometric data on V838~Mon 
with standard supergiant spectra we have derived principal stellar parameters, 
i.e. effective temperature, 
radius and luminosity, and followed the evolution of the object
since its discovery in early January~2002. Our analysis
shows that the 2002 outburst of V838~Mon consisted of two major phases: 
{\it pre-eruption}, which was 
observed in January~2002 and a major outburst, the {\it eruption}, which started
in the beginning of February~2002. During {\it pre-eruption} the object seemed to be
relaxing after an initial event which had presumably taken place in last days of
December~2001. The {\it eruption} phase, which lasted untill mid-April~2002, resulted
from a very strong energy burst, which presumably took place in last days of January
at the base of the stellar envelope inflated in {\it pre-eruption}. The burst produced
an energy wave, which was observed as a strong luminosity flash in the beginning of 
February, followed by a strong mass outflow in the form of two shells, which was observed 
as an expanding photosphere in later epochs. In mid-April, when the outflow became
optically transparent and most of its energy radiated away, the object entered the
{\it decline} phase during which V838~Mon was evolving along the Hayashi track. This
we interpret as evidence that the main energy source during 
{\it decline} was gravitational contraction of the object envelope inflated
in {\it eruption}. Late in 2002, dust formation started in the expanding shells which
gave rise to a strong infrared excess observed in 2003.

   }

   \titlerunning{Evolution of V838 Mon}

   \maketitle

\section{Introduction \label{int}}

The eruption of V838~Mon was discovered in the beginning of January~2002 
(Brown \cite{brown}). It appeared to be rather a pre-eruption
brightening as the main eruption started at the beginning of February~2002
and lasted for about two months (e.g. Munari et~al. \cite{munari}). The rise
to the main maximum in February~2002 caused to a light echo
(Henden et al. \cite{henden}) whose spectacular images were obtained
with HST (Bond et~al. \cite{bond}). As shown in Tylenda (\cite{tyl}) an
analysis of the echo images allows us to study the details of the dust
distribution near V838~Mon, as well as to constrain the distance to the
object.

Early classifications of the eruption of V838~Mon as nova-like quickly 
appeared to be inappriopriate (e.g. Munari et~al. \cite{munari}, Soker \& Tylenda
\cite{soktyl}).
The light curve of V838~Mon was atypical for novae. Also,
the observed mass loss seemed to be not enough violent
for a nova. But the strongest argument against nova-like
character came from the spectral evolution of V838~Mon. Classical novae
usually reach a minimum effective temperature near their optical maximum
brightness and then tend to evolve toward higher effective
temperature, reaching values well above $10^5$~K. V838~Mon, on the contrary,
reached the maximum effective temperature, characteristic of an A-F spectral
type, at the optical maximum in the beginning of February.
The object then
showed a general tendency to evolve to lower effective
temperatures. In mid-April~2002 it almost disappeared from optical images 
but remained
very bright in infrared, becoming one of the coolest M-type supergiants yet
observed in astrophysics. Detailed descriptions of the spectral and
photometric evolution of V838~Mon can be found in a number of papers
including Munari et~al. (\cite{munari}), Kimeswenger et~al. (\cite{kimes}),
Kolev et~al. (\cite{kolev}),
Osiwa\l{}a et~al. (\cite{osiwal}), Wisniewski et~al. (\cite{wisnia}), Crause
et~al. (\cite{lisa03}, \cite{lisa04}), Kipper et~al. (\cite{kipper}).

The global fading of V838~Mon in optical after eruption enabled us
to discover a faint hot continuum in short wavelengths (Desidera \& Munari
\cite{desmun}; Wagner \& Starrfield \cite{wagner}) later classified as a
normal B3\,V star (Munari et al. \cite{mundes}). A detailed analysis of observational
data available for the progenitor and the enviroment of V838~Mon has been
carried out in Tylenda, Soker \& Szczerba (\cite{tss}). The main conclusion of this
study is that  V838~Mon is most likely a young binary system consisting of
two intermediate mass stars. One component is the B-type companion discovered
by Munari et~al. (\cite{mundes}). 
The other component that erupted in the 2002 event has a mass
between 5--10~$M_{\sun}$ and before eruption was either a main sequence
star of similar mass and spectral type as the B-type companion or a slightly
less massive pre-main-sequence star. (However there is 
no direct evidence that both components form a real binary.)
The system is embedded in the interstellar
medium, which is most likely to be related to molecular clouds seen around the position
of V838~Mon, rather than to be expelled by the object itself as argued by 
van Loon et~al. (\cite{lers}).

The aim of this paper is to study the evolution of the effective temperature
and luminosity of V838~Mon during and after its eruption. This sort of study
is crucial for testing possible eruption mechanisms and can be an important
step towards a final understanding of the nature of V838~Mon. Our
analysis is based on photometric results published in the literature.
This is the only kind of observational data for V838~Mon where the available
results are sufficiently rich to undertake a uniform and meaningful analysis
over a time span since the discovery of the eruption untill the present
(December~2004) state of the object.

\section{Observational data and analysis  \label{obsdat}}

The analysis done in the present study is based on a comparison of
the observed photometric magnitudes to a set of reference spectra.
The least squares method has been used to get the best fit between the model spectrum
and the observations. Since V838~Mon is considerably reddened (see Sect.~\ref{ext_d})
the model spectra have been reddened before fitting them to the observations.
The standard extinction curve, $R \equiv A_V/E_{B-V} = 3.1$, has been adopted.
From the best fit one obtains an effective
temperature, $T_\mathrm{eff}$, and an angular stellar radius, $\theta$. 
If the distance to the object is known, a linear radius, $R$, and a stellar 
luminosity, $L$, can then be calculated.

It is clear that the larger the wavelength range that is covered by the photometric
measurements, the more reliable are the results obtained.
Therefore we have primarily based our analysis on $UBVR_cI_cJHKL$ magnitudes
available from Munari et~al. (\cite{munari}) and Crause et~al. (\cite{lisa03},
\cite{lisa04}). For certain time
periods no $JHKL$ measurements are available and we had
to rely on $UBVR_cI_c$ data from the above references, supplemented
by the $BVRI_c$ measurements of Kimeswenger et~al. (\cite{kimes}).
In later epochs when the object became very cool, the infrared data were
particularly important. Therefore we have also
used photometric data from Henden \& Munari (\cite{henmun}),
Watson \& Costero (\cite{watcos}), Lynch et~al.
(\cite{lynch03}, \cite{lynch04}) and Tapia \& Persi (\cite{tapia}).

Given the luminosity range obtained in the present study 
($10^4 - 10^6~L_{\sun}$) as well as the spectral appearance of V838~Mon
during eruption (e.g. Kipper et~al. \cite{kipper}) we have used standard
intrinsic colours for supergiants (luminosty class I) when constructing 
the set of reference spectra. The intrinsic photometric colours have been
taken from Schmidt-Kaler (\cite{sk} -- $UBV$), Johnson (\cite{johnson}
-- $VRI$ for FGK types), Lee (\cite{lee} -- $VRI$ for M types)
and Koornneef (\cite{koor} -- $VJHKL$). A calibration of the effective
temperature and bolometric correction against the spectral type has been
taken from Schmidt-Kaler (\cite{sk}). 

In this way we obtained a grid of 21
reference spectra ranging from F0\,I to M5\,I ($T_\mathrm{eff}$ ranging
between 7700 and 2800~K). Between the grid points a linear interpolation 
was done either
in the colour -- log~$T_\mathrm{eff}$ plane (e.g. $B-V$ or $V-K$ versus
log~$T_\mathrm{eff}$) to get a reference spectrum
for any given value of $T_\mathrm{eff}$ or in the colour -- log~$\lambda$ plane
if the effective wavelength of a photometric band (e.g. $R_c$, $I_c$) differed
from the Johonson standards.

After mid-April 2002 V838~Mon became cooler than the M5\,I standard. In order
to have estimates of the object parameters in later epochs we proceeded in 
two ways. First, we extrapolated our grid of standard spectra down to M10. 
The extrapolation is in the log~colour -- log~$T_\mathrm{eff}$ plane
(including log~BC -- log~$T_\mathrm{eff}$). In the range of M types we explore
the Wien part of the spectrum and consequently the relations are roughly
linear in the above plane. Second, we fitted a blackbody spectrum to the
observations. For the
effective wavelengths of particular photomteric bands, the fluxes
obtained from the blackbody law have been converted to magnitudes using the
standard calibration of Vega (Schmidt-Kaler \cite{sk}; Tokunaga
\cite{tokun}). After having reddened them according to 
the extinction curve, the magnitudes were compared to
the observations.

\section{Distance and interstellar extinction \label{ext_d}}

Initial estimates of the distance to V838~Mon done in Munari et~al.
(\cite{munari}) and Kimeswenger et~al. \cite{kimes}) gave values of 
0.6--0.8~kpc. They were however based on a wrong interpretation of the
observed expansion of the light echo. From a more realistic study of the echo
structure Bond et~al. (\cite{bond}) concluded that the distance is at least
6~kpc. A detailed study of the light echo
evolution done in Tylenda (\cite{tyl}) and Tylenda et~al. (\cite{tss}) 
shows that the distance cannot be
unambiguously determined from the observed expansion 
of the outer echo edge. Only a lower limit of $\sim 5$~kpc can be established.
However, from an analysis of the inner structure of the echo Tylenda (\cite{tyl})
estimated that V838~Mon is at a distance of $8.0\pm2.0$~kpc. 
More recently, Crause et~al. (\cite{lisa04}) made an analysis of the echo
expansion observed at the SAAO between May~2002 and December~2004 and conclude that
the distance is $\sim 9$~kpc.

Wisniewski et~al. (\cite{wisnia}) have analysed radial velocity
structures in the interstellar NaI~D lines observed in the spectrum of V838~Mon. 
From a comparison with a velocity contour map in the Galaxy they conclude 
that the lower limit on the distance is $\sim 2.5$~kpc. From observations
of the same lines Kipper et~al. (\cite{kipper}) determined a reddening
distance $>3$~kpc and a kinematic distance $>4$~kpc.

The identification of the B-type companion in the spectrum of V838~Mon 
in Munari et~al. (\cite{mundes}) enables
the application of the spectroscopic parallax method. 
Adopting the photomatric data, spectral classification and $E_{B-V}$ 
from Munari et~al (\cite{mundes}) as well as
the absolute magnitude $M_V = -1.6$ for the spectral class B3\,V 
from Schmidt-Kaler (\cite{sk}) one gets a distance
to the object of 9.4~kpc. (Munari et~al., \cite{mundes}, give d = 10.5~kpc, 
presumably adopting
a somewhat different calibration of $M_V$.) A lower limit to the distance
from the above data can be obtained adopting the ZAMS calibration of Schmidt-Kaler 
(\cite{sk}) which gives $M_V = -1.1$ for the B3\,V type. The distance then decreases
to 7.5~kpc.
From the photometric data of Crause et~al. (\cite{lisa04}),
Tylenda et~al. (\cite{tss}) derived parameters for the B-type companion
slightly different from those of Munari et~al. (\cite{mundes}), 
namely $V = 16.26 \pm 0.02$, $E_{B-V} = 0.71$ and a B4\,V spectral type.
In this case one obtaines a distance of 12.4~kpc 
and 9.4~kpc using the main sequence and ZAMS calibrations of Schmidt-Kaler 
(\cite{sk}), respectively.

The conclusion of Tylenda et~al. (\cite{tss}) that V838~Mon is very likely
to be a young binary system allows us to assume that the system follows more or less
the Galactic disc rotation and to estimate the distance from the radial velocity
of the system. Kolev et~al. (\cite{kolev})
from the fairly symmetric emission H$\alpha$ profile in January~2002 have derived
a heliocentric wavelength of the line of 6564.155\,\AA\ which results in 
a heliocentric radial velocity of +62~$\mbox{km}\ \mbox{s}^{-1}$. Kipper et~al. (\cite{kipper}) obtained
$+59\pm6\ \mbox{km}\ \mbox{s}^{-1}$ from emission line profiles in February~2002. The line profiles
were rather complex so this estimate is uncertain. T.~Tomov (2004, private communication)
derived a radial velocity of $\sim +64\ \mbox{km}\ \mbox{s}^{-1}$ from 
the H$\beta$ absorption
line in the spectrum of the B-type component. The
above values agree very well with the radial velocity of a more redshifted 
component in the interstellar NaI~D lines, i.e. $+64\pm1\ \mbox{km}\ \mbox{s}^{-1}$ 
(Zwitter \& Munari \cite{zwimun}, Kolev et~al. \cite{kolev}, Kipper et~al. \cite{kipper}).
This supports the conclusion of Tylenda et~al. (\cite{tss}) that V838~Mon is
embedded in the interstellar medium. Thus we can adopt that the heliocentric
radial velocity of V838~Mon is +64~$\mbox{km}\ \mbox{s}^{-1}$. Using the results of
Dehnen \& Binney (\cite{debin}) this can be transformed to 
$V_\mathrm{LSR} = +53\ \mbox{km}\ \mbox{s}^{-1}$ which, 
adopting the Galactic rotation curve of
Brand \& Blitz (\cite{bb}), gives a distance of $\sim 6.7$~kpc.

Munari et~al. (\cite{mun05}) from an analysis of interstellar
extinction, galactic kinematics and properties of the B-type companion 
have concluded that the distance to V838~Mon is $\sim 10$~kpc.

Summarizing the above distance determinations we can conclude that observations 
of V838~Mon
itself and of its B-type companion lead to consistent results
and that the system is most likely at a distance between 5 and 12~kpc.
For the purpose of the present study we assume that the distance to V838~Mon 
is 8.0~kpc. 

From interstellar components of K~I and Na~I lines in the spectrum of V838~Mon
Zwitter \& Munari (\cite{zwimun}) estimated an interstellar extinction
$E_{B-V} = 0.80 \pm 0.05$ and concluded that this value would imply 
a distance $\ga 3$~kpc. Munari et~al. (\cite{munari}) have, however,
argued that this value of extinction was too large as their distance 
estimate gave 790~pc and they adopted $E_{B-V} = 0.50$. 
This distance estimate was however wrong, as discussed above, 
and V838~Mon is certainly more distant than
3~kpc. Therefore the $E_{B-V}$ estimate in Zwitter \& Munari
(\cite{zwimun}) can be considered as reasonable. 

Kimeswenger et~al. (\cite{kimes}) have compared effective temperatures estimated from
spectroscopy with photometric colours and found that $0.5 \la E_{B-V} \la 0.9$.
For their final discussion Kimeswenger et~al. adopted $E_{B-V} \simeq 0.7$.

Kipper et~al. (\cite{kipper}) have applied different methods for estimating
the reddening. From equivalent widths of the Na~I interstellar lines
they derived $E_{B-V} = 0.49$. Using equivalent widths
of diffuse interstellar bands they obtained $E_{B-V}$ between 0.41 and 2.1.
From dust distribution maps these authors estimated an upper limit
of $E_{B-V} \simeq 0.89$ in the direction of V838~Mon.

Probably the most reliable estimate of the reddening done in Kipper et~al.
is that resulting from their Table~3. These authors have classified their
spectra of V838~Mon obtained between 4~February and 2~April~2002 using
comparison spectra of supergiants. The $(B-V)$ values expected from 
the classification are then compared with the observed values giving
$E_{B-V}$. From our compilations we have supplemented the missing data 
in their Table~3 for 26~March and 2~April and then from all the results 
we have obtained $E_{B-V} = 1.09 \pm 0.17$.

As discussed in Tylenda et~al. (\cite{tss}) the available photometric data on 
the B-type companion give $E_{B-V}$ between 0.7 and 1.0.

In a recent paper Munari et~al. (\cite{mun05}) have determined the interstellar
extinction from interstellar NaI and KI lines, distribution of the extinction
along the line of sight and properties of the B-type companion. They conclude
that $E_{B-V} \simeq 0.87$.

In the present study we assume $E_{B-V} = 0.9$. This value is 
a good compromise between the various results obtained for V838~Mon itself
and its B-type companion, especially if estimates as low as 0.5 
are not considered as being too low for a distance $>5$~kpc near the
Galactic plane. We also note that the best fits between the photometric 
results analysed in the present study and the reference supergiant spectra
are in most cases obtained when $E_{B-V} \simeq 0.9$ is adopted.

\section{Results  \label{results}}

Throughout the present paper the epoch of observations is given in days
counted since 1~January~2002. We devide the time span discussed 
into three phases (for the light curve of V838~Mon see e.g. Munari et~al.
\cite{munhen}, \cite{munari} or Crause et~al. \cite{lisa03}, \cite{lisa04}). 
The first one, called {\it pre-eruption}, lasted untill 
1~February~2002, i.e. it is includes epochs $\le 32$~days. The object then slowly
evolved between $V = 10 - 11$. The second one, called {\it eruption}, lasted untill 
mid-April~2002 covering the epochs between 32 and $\sim 100$~days. The object
was then brighter than 9~magnitudes, twice reaching $V \simeq 7$.
In mid-April V838~Mon started a fast decline in $V$, the period with
epochs $\ga\ $100~days is called {\it decline}.

\begin{figure*}
  \centering
  \includegraphics[width=8.9cm]{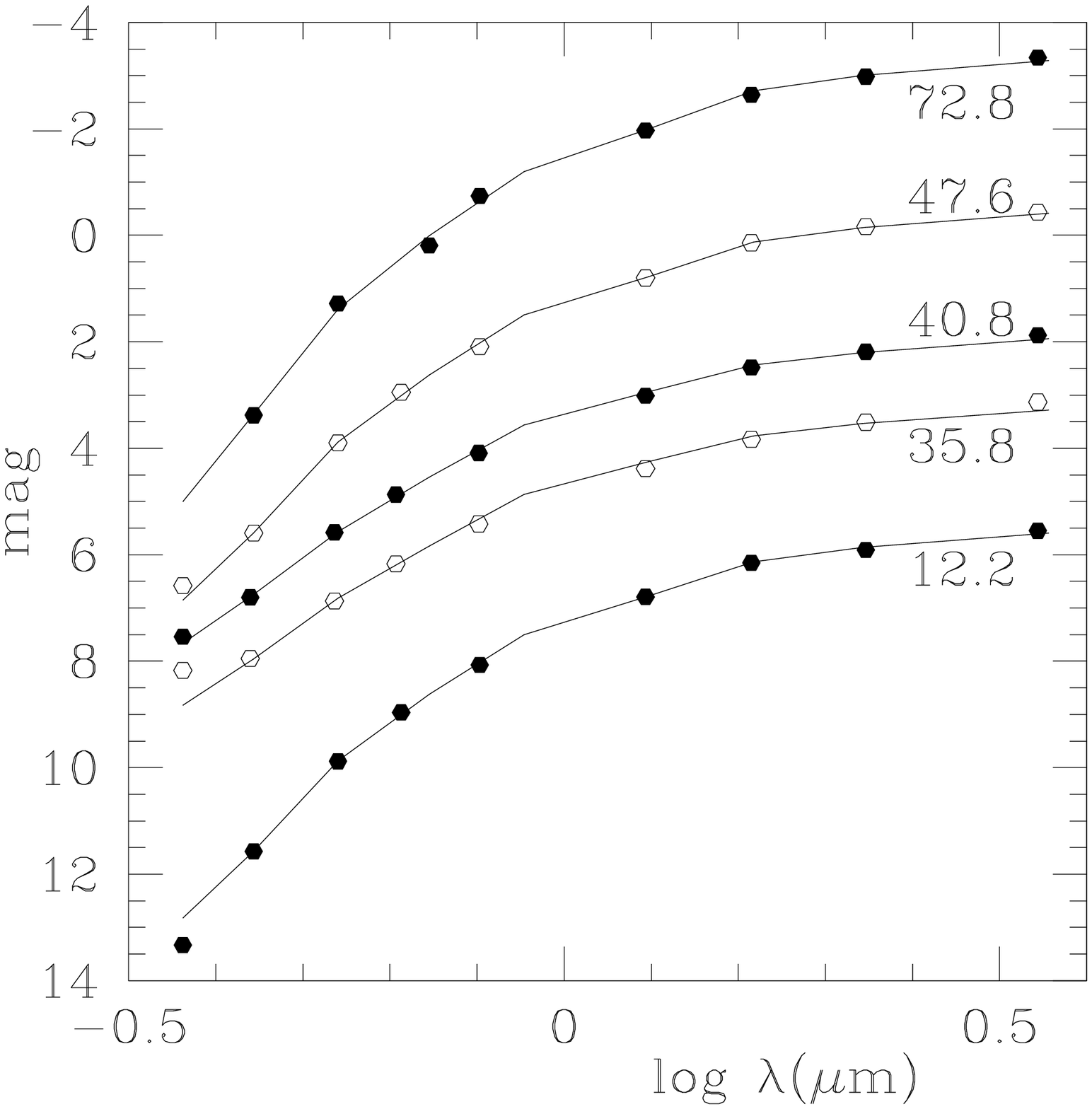}
  \includegraphics[width=8.9cm]{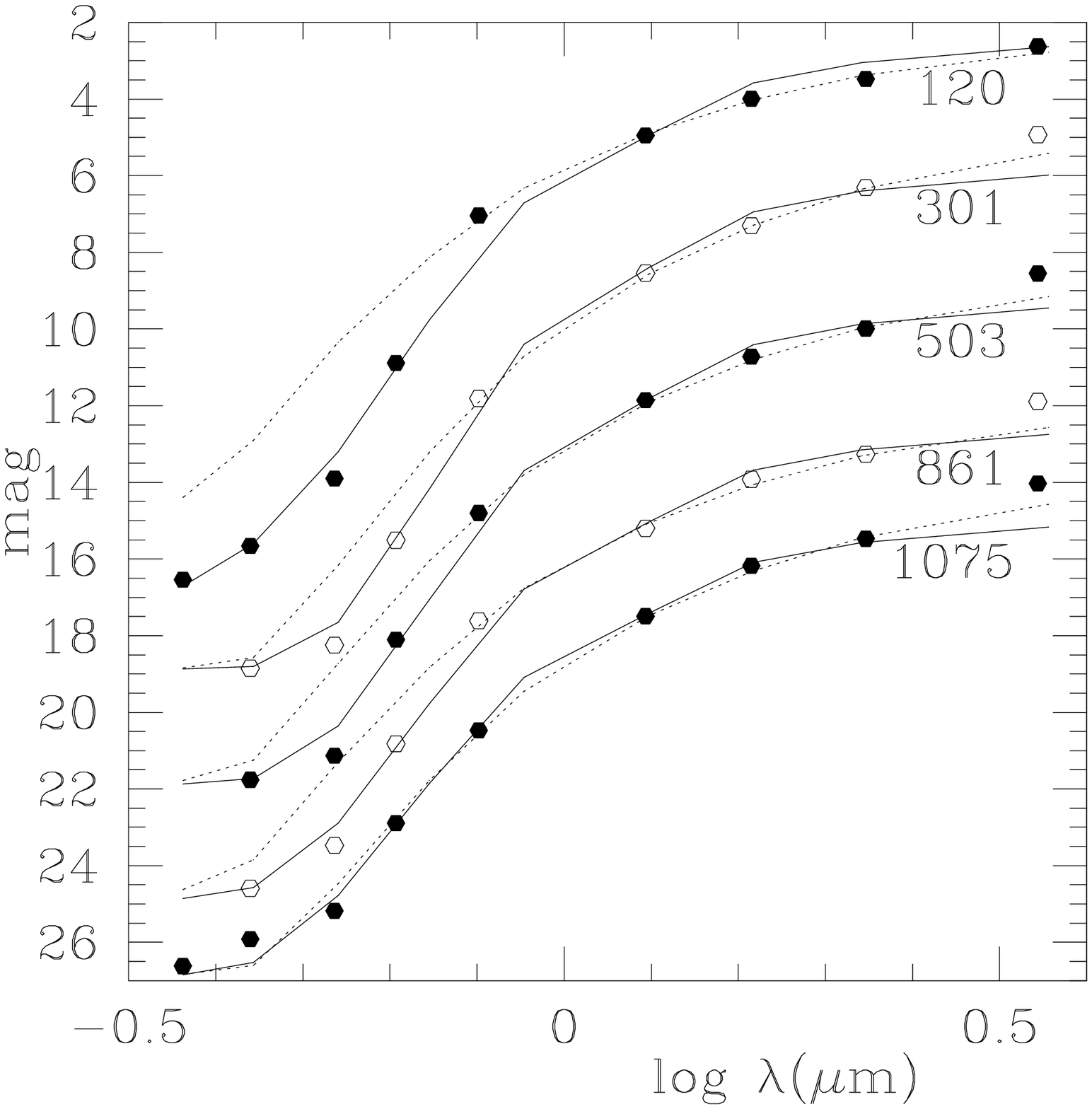}
  \caption{Examples of the fits of reference spectra 
to the observed magnitudes of V838~Mon. 
{\bf Left panel}: {\it pre-eruption} and
{\it eruption} phases. {\bf Right panel}: {\it decline} phase. 
Symbols -- observed magnitudes. 
Full curves -- supergiant spectra. Dotted curves - blackbody distributions.
All reference spectra have been reddened with $E_{B-V} = 0.9$.
The curves are labelled with the epoch of observations (days
since 1~Jan.~2002). Note that
the curves and the data corresponding to days 40.8, 47.6
and 72.8 (left panel) have been shifted by $-$2, $-$4 and $-$6 magnitudes, 
respectively, while those corresponding to days 301, 503,
861 and 1075 (right panel) 
have been shifted by 2, 5, 8 and 10 magnitudes, respectively.
Parameters of the fits can be found in Table~\ref{evol_t}.}
  \label{fig_s}
\end{figure*}

Figure~\ref{fig_s} presents examples of the reference spectra 
(full curves: supergiants -- dotted curves: blackbodies)
fitted to the photometric data (symbols).
The individual spectra have been labelled with the epoch of observations.

Table~\ref{evol_t} shows the results from fitting the photometric data. Column (1)
shows the epoch of observations (days since 1~January~2002). Column (2) gives
the spectral range used in the fitting procedure (e.g. $BI$ stands for $BVRI$ while
$BL$ means $BVRIJHKL$) and the references to the data
(explained in the bottom of Table~\ref{evol_t}). The spectral type,
the corresponding effective temperature and the angular photospheric radius
($\theta$ is in radians) resulting from
fitting the standard supergiant spectra to the data are presented in columns (3), (4)
and (5), respectively. The (linear) photospheric radius and the luminosity, given
in columns (6) and (7), have been calculated assuming a distance of 8~kpc.
The second line, labelled BB in 
column (3), for epochs greater than 110~days gives
the results of the fits using blackbodies as reference spectra.

\begin{table}[t]
\centering
\caption{Results from fitting the supergiant and blackbody 
spectra (the latter only for day $>$110) to the photometric data. 
See text for explanations of the columns.}
\label{evol_t}
\begin{tabular}{@{} r l l r @{~~~} r @{~~~} r @{~~~} r @{}}
\hline
   day & data & Sp. & $T_\mathrm{eff}$ & $-$lg\,$\theta$ & $R/R_{\sun}$
& $L/L_{\sun}$ \\
\hline
   10.1 & $BL$--m2 & G2.1 &  5190.  &  9.00 &  360. &  8.3E4 \\
   12.2 & $BL$--c3,m2 & G1.0 &  5370.  &  9.01 &  340. &  8.8E4 \\
   13.3 & $BI$--m2 & G0.9 &  5390. &  9.02 &  340. &  8.7E4 \\
   13.9 & $BI$--k2 & G1.0 &  5370. &  9.02 &  340. &  8.5E4 \\
   14.9 & $BI$--k2 & G1.5 &  5280. &  9.02 &  340. &  8.0E4 \\
   14.9 & $BI$--c3 & G1.0 &  5370. &  9.02 &  340. &  8.8E4 \\
   16.8 & $BI$--k2 & G2.4 &  5140. &  9.01 &  350. &  7.7E4 \\
   18.9 & $BI$--c3 & G2.9 &  5080. &  8.99 &  360. &  7.8E4 \\
   19.9 & $BI$--c3 & G2.5 &  5130. &  9.00 &  350. &  7.8E4 \\
   24.8 & $BI$--k2 & G5.6 &  4810. &  8.98 &  370. &  6.5E4 \\
   28.9 & $BI$--k2 & G7.1 &  4680. &  9.00 &  360. &  5.5E4 \\
   31.8 & $BI$--k2 & G8.0 &  4600. &  9.02 &  340. &  4.7E4 \\
   32.8 & $BI$--k2 & F5.8 &  6660. &  8.90 &  440. &  3.5E5 \\
   33.0 & $BI$--k2 & F5.4 &  6790. &  8.88 &  470. &  4.2E5 \\
   33.8 & $BI$--k2 & F4.7 &  6950. &  8.81 &  550. &  6.4E5 \\
   34.1 & $BL$--m2 & F5.0 &  6900. &  8.78 &  590. &  7.1E5 \\
   34.9 & $BI$--k2 & F2.7 &  7240. &  8.77 &  600. &  9.0E5 \\
   35.8 & $BL$--c3,m2 & F2.7 &  7240. &  8.70 &  710. &  1.25E6 \\
   37.8 & $BL$--c3,m2 & F3.2 &  7170. &  8.72 &  680. &  1.10E6 \\
   38.8 & $BL$--c3,m2 & F3.2 &  7170. &  8.76 &  610. &  9.0E5 \\
   38.9 & $BI$--k2 & F2.2 &  7310. &  8.79 &  580. &  8.5E5 \\
   39.2 & $BI$--m2 & F2.2 &  7310. &  8.80 &  570. &  8.3E5 \\
   39.8 & $BL$--c3,m2 & F3.2 &  7160. &  8.80 &  560. &  7.4E5 \\
   40.1 & $BI$--m2 & F3.1 &  7170. &  8.81 &  550. &  7.2E5 \\
   40.8 & $BL$--c3,m2 & F4.1 &  7040. &  8.82 &  540. &  6.4E5 \\
   41.2 & $BI$--m2 & F3.1 &  7170. &  8.85 &  500. &  6.0E5 \\
   41.8 & $BI$--k2 & F5.9 &  6660. &  8.80 &  570. &  5.7E5 \\
   42.0 & $BL$--m2 & F5.8 &  6660. &  8.80 &  560. &  5.6E5 \\
   42.8 & $BI$--k2 & F6.7 &  6420. &  8.78 &  590. &  5.3E5 \\
   42.9 & $BL$--c3,m2 & F6.5 &  6480. &  8.78 &  590. &  5.5E5 \\
   43.9 & $BL$--c3,m2 & F7.5 &  6220. &  8.75 &  640. &  5.5E5 \\
   44.9 & $BL$--c3,m2 & F8.6 &  5930. &  8.70 &  710. &  5.5E5 \\
   45.9 & $BL$--c3,m2 & F9.3 &  5720. &  8.68 &  750. &  5.4E5 \\
   45.9 & $BI$--k2 & F9.6 &  5660. &  8.70 &  750. &  5.2E5 \\
   46.8 & $BI$--k2 & G0.2 &  5510. &  8.65 &  800. &  5.4E5 \\
   47.6 & $BL$--m2 & G1.2 &  5330. &  8.61 &  880. &  5.6E5 \\
   47.8 & $BI$--k2 & G0.3 &  5500. &  8.65 &  800. &  5.2E5 \\
   48.7 & $BL$--c3,m2 & G2.5 &  5130. &  8.57 &  960. &  5.7E5 \\
   50.9 & $BI$--c3 & G2.8 &  5090. &  8.57 &  960. &  5.6E5 \\
   51.8 & $BI$--c3 & G3.8 &  4970. &  8.54 & 1010. &  5.6E5 \\
   55.2 & $BI$--m2 & G8.9 &  4520. &  8.45 & 1260. &  6.0E5 \\
   55.8 & $BI$--c3 & G8.4 &  4560. &  8.46 & 1230. &  5.9E5 \\
   57.1 & $BI$--m2 & K0.6 &  4360. &  8.42 & 1360. &  6.0E5 \\
   58.1 & $BI$--m2 & K0.6 &  4360. &  8.41 & 1380. &  6.2E5 \\
   58.8 & $BI$--k2 & G9.3 &  4480. &  8.44 & 1290. &  6.0E5 \\
   59.8 & $BI$--c3 & G6.3 &  4740. &  8.47 & 1200. &  6.6E5 \\
   60.8 & $BI$--c3 & G4.4 &  4920. &  8.49 & 1140. &  6.9E5 \\
   62.2 & $BI$--m2 & G3.6 &  5000. &  8.48 & 1170. &  7.7E5 \\
   62.9 & $BI$--k2 & G3.4 &  5020. &  8.48 & 1180. &  8.0E5 \\
   63.2 & $BI$--m2 & G2.1 &  5180. &  8.50 & 1120. &  8.2E5 \\
\hline
\end{tabular}
\end{table}
\begin{table}[t]
\centering
\centerline{Table \ref{evol_t} (continued)}
\centerline{}
\begin{tabular}{@{} r l l r @{~~~} r @{~~~} r @{~~~} r @{}}
\hline
   day & data & Sp. & $T_\mathrm{eff}$ & $-$lg\,$\theta$ & $R/R_{\sun}$
& $L/L_{\sun}$ \\
\hline
\hline
   63.8 & $BI$--k2 & G3.0 &  5060. &  8.47 & 1210. &  8.6E5 \\
   65.8 & $BI$--k2 & G2.5 &  5130. &  8.45 & 1260. &  9.9E5 \\
   66.8 & $BI$--k2 & G2.5 &  5130. &  8.44 & 1290. &  1.05E6 \\
   67.1 & $BI$--m2 & G2.7 &  5100. &  8.42 & 1340. &  1.10E6 \\
   67.8 & $BI$--k2 & G2.6 &  5110. &  8.43 & 1320. &  1.07E6 \\
   68.2 & $BI$--m2 & G2.7 &  5100. &  8.42 & 1340. &  1.10E6 \\
   68.8 & $BI$--k2 & G2.9 &  5070. &  8.42 & 1350. &  1.08E6 \\
   69.1 & $BI$--m2 & G2.4 &  5140. &  8.42 & 1330. &  1.12E6 \\
   69.8 & $BI$--k2 & G3.4 &  5010. &  8.41 & 1370. &  1.07E6 \\
   71.8 & $BL$--c3,k2 & G5.3 &  4830. &  8.38 & 1460. &  1.05E6 \\
   72.8 & $BL$--c3,k2 & G6.6 &  4710. &  8.37 & 1520. &  1.03E6 \\
   73.8 & $BI$--k2 & G8.5 &  4560. &  8.33 & 1650. &  1.06E6 \\
   85.8 & $BI$--k2 & K3.0 &  4090. &  8.23 & 2080. &  1.09E6 \\
   86.8 & $BI$--k2 & K3.2 &  4060. &  8.22 & 2120. &  1.10E6 \\
   88.8 & $BI$--k2 & K3.7 &  4000. &  8.21 & 2200. &  1.11E6 \\
   90.2 & $BI$--m2 & K4.1 &  3950. &  8.18 & 2350. &  1.21E6 \\
   91.8 & $BI$--k2 & K4.3 &  3920. &  8.20 & 2250. &  1.07E6 \\
   93.8 & $BI$--k2 & K4.4 &  3910. &  8.20 & 2250. &  1.07E6 \\
   98.1 & $BI$--m2 & K9.8 &  3640. &  8.14 & 2580. &  1.05E6 \\
   99.8 & $BI$--k2 & K9.1 &  3670. &  8.18 & 2340. &  8.9E5 \\
  102.8 & $BI$--k2 & M1.0 &  3550. &  8.20 & 2250. &  7.2E5 \\
  119.8 & $BL$--c3 & M6.7 &  2490. &  8.10 & 2800. &  2.7E5 \\
        & $IL$     & BB   &  3060. &  8.30 & 1760. &  2.5E5 \\
  120.7 & $BL$--c3 & M7.1 &  2420. &  8.08 & 2920. &  2.6E5 \\
        & $IL$     & BB   &  3000. &  8.30 & 1790. &  2.3E5 \\
  153.2 & $BL$--c3 & M9.4 &  2020. &  8.08 & 2970. &  1.3E5 \\
        & $IL$     & BB   &  2070. &  8.16 & 2430. &  9.8E4 \\
  232.0 & $BL$--hm & M11  &  1780. &  8.05 & 3190. &  9.2E4 \\
        & $IL$     & BB   &  1710. &  8.06 & 3120. &  7.5E4 \\
  254.1 & $BL$--c5 & M9.7 &  1980. &  8.16 & 2450. &  8.3E4 \\
        & $IL$     & BB   &  1840. &  8.14 & 2540. &  6.7E4 \\
  301.0 & $BK$--c5 & M8.5 &  2180. &  8.26 & 1960. &  7.8E4 \\
        & $IK$     & BB   &  2120. &  8.27 & 1890. &  6.5E4 \\
  385.9 & $BK$--c5 & M8.2 &  2230. &  8.32 & 1690. &  6.4E4 \\
        & $IK$     & BB   &  2210. &  8.36 & 1560. &  5.2E4 \\
  391.8 & $BK$--c5 & M8.1 &  2240. &  8.33 & 1670. &  6.3E4 \\
        & $IK$     & BB   &  2220. &  8.37 & 1530. &  5.1E4 \\
  474.0 & $BK$--c5 & M7.6 &  2330. &  8.40 & 1410. &  5.3E4 \\
        & $IK$     & BB   &  2390. &  8.47 & 1210. &  4.3E4 \\
  503.0 & $BK$--c5 & M7.5 &  2340. &  8.41 & 1370. &  5.0E4 \\
        & $IK$     & BB   &  2380. &  8.47 & 1190. &  4.1E4 \\
  619.1 & $BK$--c5 & M6.8 &  2470. &  8.48 & 1180. &  4.7E4 \\
        & $IK$     & BB   &  2580. &  8.55 & 1000. &  4.0E4 \\
  650.0 & $BK$--c5 & M6.8 &  2480. &  8.49 & 1140. &  4.4E4 \\
        & $IK$     & BB   &  2520. &  8.54 & 1020. &  3.8E4 \\
  861.0 & $BK$--c5 & M6.6 &  2510. &  8.53 & 1050. &  3.9E4 \\
        & $IK$     & BB   &  2630. &  8.60 &  890. &  3.4E4 \\
 1075.0 & $BK$--c5 & M5.9 &  2620. &  8.64 &  800. &  2.8E4 \\
        & $IK$     & BB   &  2280. &  8.54 & 1020. &  2.6E4 \\
\hline
\end{tabular}
\\
\begin{flushleft}
Data references: 
c3 -- Crause et~al. (\cite{lisa03}),
c5 -- Crause et~al. (\cite{lisa04}),
hm -- Henden \& Munari (\cite{henmun}),
k2 -- Kimeswenger et~al. (\cite{kimes}),
m2 -- Munari et~al. (\cite{munari})
\end{flushleft}
\end{table}

The left panel of Fig.~\ref{fig_s} displays the spectra during the {\it pre-eruption}
and {\it eruption} phases.
The $U$ magnitude has not been taken into account in the fitting 
procedure. This band is dominated by the Balmer continuum which is very sensitive to
nonstandard phenomena such as departures from hydrostatic equilibrium, non-LTE
and winds. As can be seen from Fig.~\ref{fig_s} (day 12.2), 
in {\it pre-eruption} the $U$ magnitude was below that expected 
for standard supergiants. However, during the sharp maximum at the begining of February
the Balmer continuum was well above that in supergiants (see day 35.8 in
Fig.~\ref{fig_s}). This period was also marked by the appearance of strong Balmer
emission lines (see e.g. Osiwa\l{}a \cite{osiwal}). 
Also the quality of the obtained
spectral fit was then less good in the sense that the observed spectrum 
was somewhat flatter than the standard one.
In later epochs, after the sharp maximum, the overall spectral fit became 
better, including 
the $U$ magnitude (see days 40.8 and 47.6 in Fig.~\ref{fig_s}). Thus the object
then resembled a standard supergiant quite closely.

As can be seen from Table~\ref{evol_t}, for a number of epochs 
in {\it pre-eruption} and {\it eruption} the fitting has been 
done using only the $BVRI$ magnitudes, as no $JHKL$ measurements were available
for these dates. The results from these fits are expected to be less reliable.
However, as discussed in Sect.~\ref{evol}, there is no significant systematic difference
between them and those based on the $BVRIJHKL$ data.

The right panel of Fig. \ref{fig_s} shows examples of the fits obtained in 
the {\it decline} phase. 
As explained in Sect.~\ref{obsdat}, the fits for this period were obtained
using both the extrapolated standard supergiant spectra and the blackbody
distributions. In all fits in the {\it decline}
period a contribution from the B-type companion, the same as
in Tylenda et~al. (\cite{tss}), has been taken into account.
This contribution is important only in the $UBV$ bands. As can be seen from 
the right panel of Fig.~\ref{fig_s}, the observed spectra change their slope 
below the $I$ band, i.e.
for shorter wavelength bands the spectrum is steeper than for the longer ones. 
This effect, due to strong molecular bands significantly lowering the flux
at shorter wavelengths, is roughly reproduced in the supergiant spectra.
The blackbody spectra do not take this into account. Therefore, when fitting 
blackbodies
we excluded the $BVR$ magnitudes from the fitting procedure.
For the epochs later than 300~days the $L$ magnitude was also omitted from the
fitting (both using supergiant as well as blackbody spectra). This
band was likely to be affected by the infrared excess discussed in Sect.~\ref{ire}.
Inclusion or exclusion of the $L$ magnitude in the fitting procedure
does not significantly change the fit results.

As can be seen from Fig.~\ref{fig_s}, given the crudeness of the reference spectra
used in the {\it decline} phase (extrapolated supergiants or blackbodies), 
the obtained fits to the observations
are reasonable (although not as good as those in {\it pre-eruption}
and {\it eruption}). The values
of the fitting parameters ($T_\mathrm{eff}$ and $\theta$ in Table~\ref{evol_t})
obtained using the supergiant spectra are not significantly different from those
using the blackbodies in most cases. Only for two spectra obtained near
day 120 does fitting a blackbody give a $T_\mathrm{eff}$ larger by almost 600~K
and $\theta$ smaller by a factor of $\sim 1.6$ compared to the results of fitting
a supergiant. As discussed in Sect.~\ref{erup}, the object was probably then 
surrounded by an extended shell in a transition from optically thick
to optically thin. In the case of the last spectrum (day 1075) the blackbody 
fitting also gave $T_\mathrm{eff}$ about 350~K lower and, consequently, a somewhat larger
effective radius than the supergiant fitting.

Note that for the reasons discussed above we do not consider fits based only on
$BVRI$ in the {\it decline} phase.

Lane et al. (\cite{lane}) attempted to determine the 
angular size of V838~Mon from interferomatric observations done in the $K$ band
in November--December~2004. Their result of an angular diameter of 1.8~mas is
slightly larger than our result obtained on day 1075 ($\sim 1.1$~mas). One possible 
reason for this difference is that the uniform disc used in Lane et~al. to deconvolve
the data might be not a good model for the photosphere of V838~Mon. 
Another possible reason is that
although the $K$ band, according to our analysis, is dominated by the central object,
a certain contribution from an extended infrared emission, discussed in Sect.~\ref{ire},
may also be important. This might explain why an elliptical model better fits
the data of Lane et~al. than a circular one. It thus would be desirable to make similar
interferometric measurements but in shorter wavelengths, say in $I$ or $J$, 
where the central object certainly dominates the flux, as well as in longer wavelengths, 
say in $M$ or $N$, where the expanding matter is dominant.

\section{Evolution of the object  \label{evol}}

\begin{figure}
\centering
  \includegraphics[width=7.4cm]{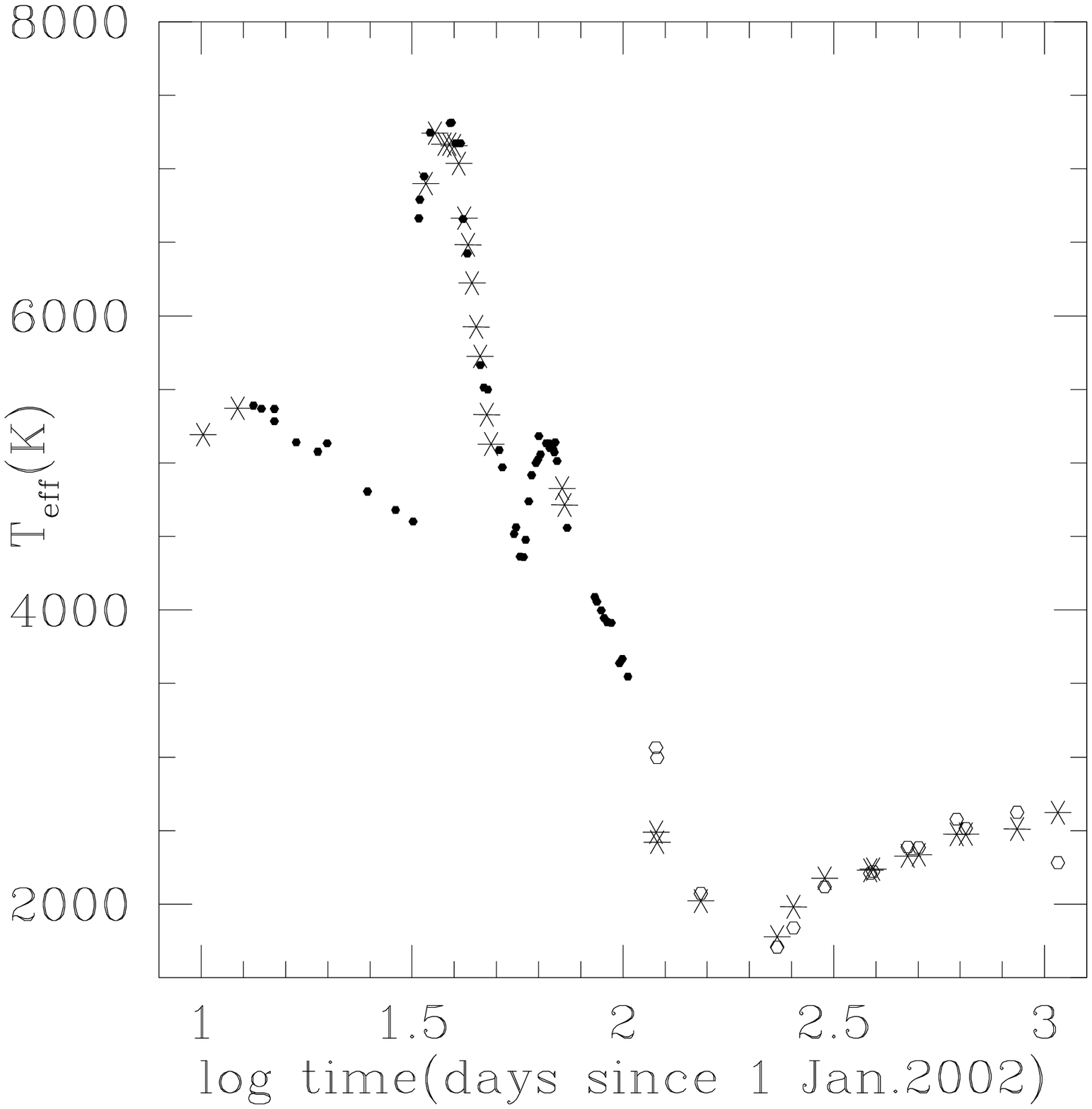}
  \includegraphics[width=7.4cm]{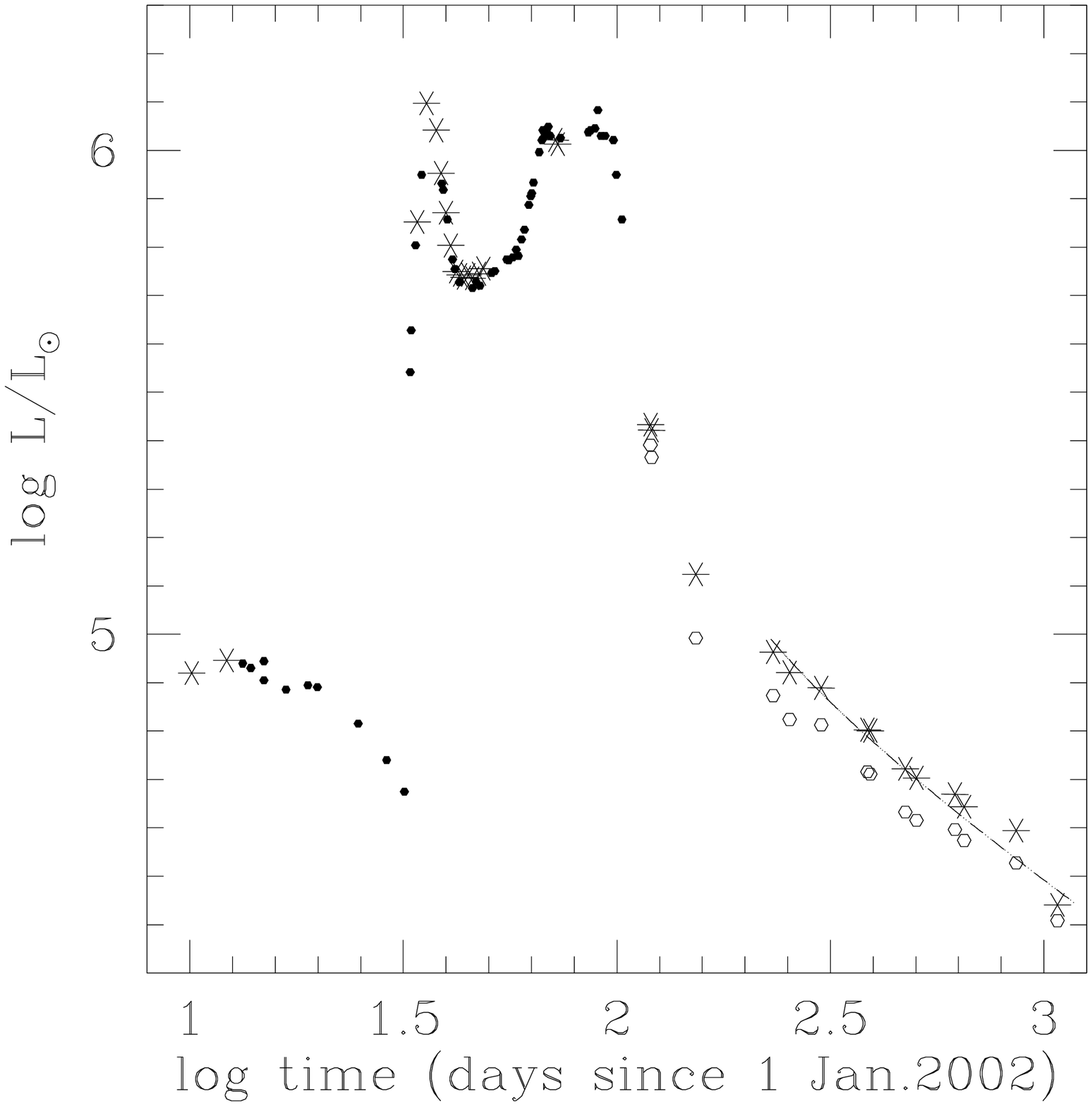}
  \includegraphics[width=7.4cm]{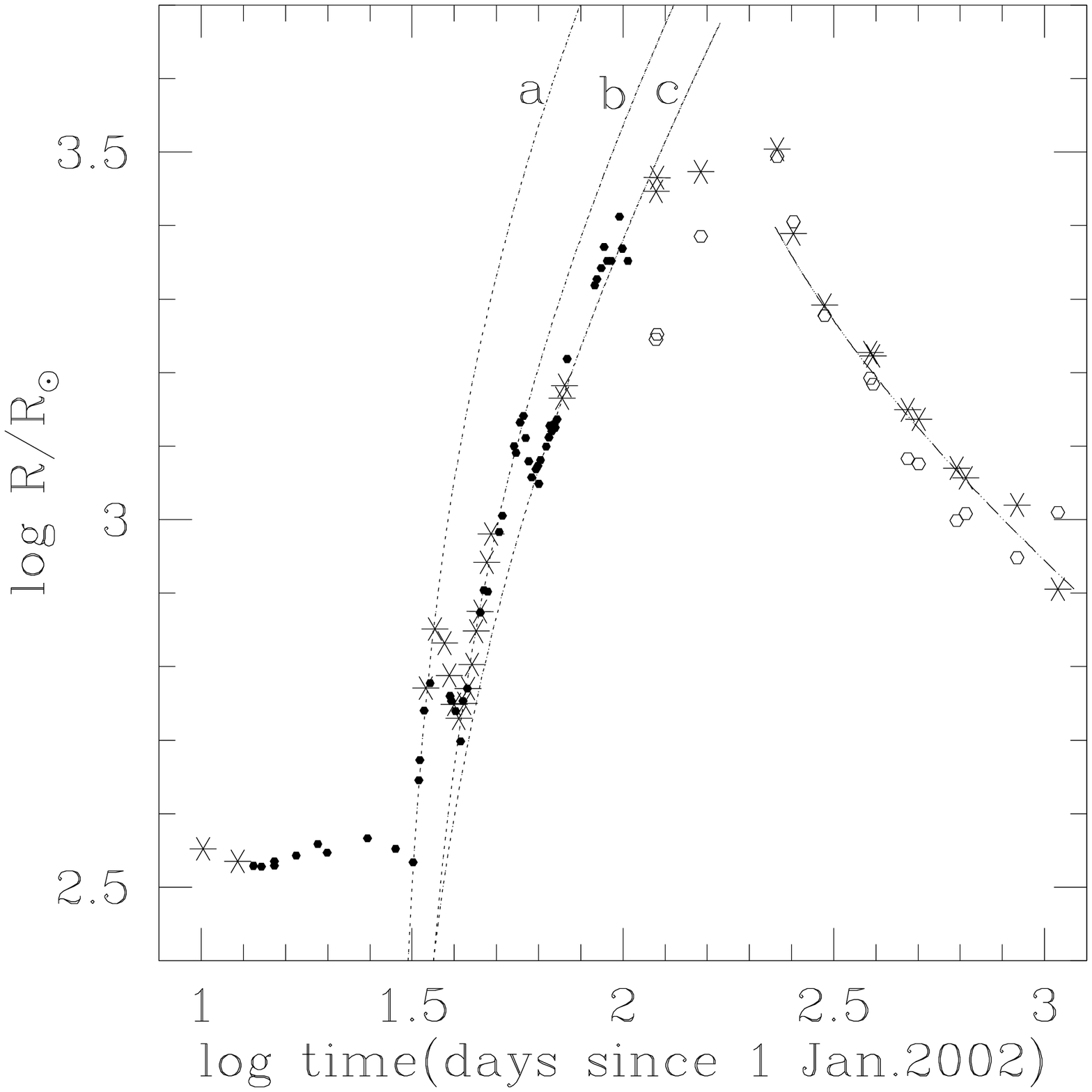}
  \caption{Evolution of the effective temperature ({\bf upper}),
luminosity ({\bf middle}) and radius ({\bf bottom}) of V838~Mon.
The luminosity and radius are in solar units. The abscissa gives the
logarithm of time in days counted since 1~Jan.~2002. See text for
explanation of the symbols and curves (in the middle and bottom panels).
}
  \label{evol_f}
\end{figure}
                                                                                    
Figure \ref{evol_f} shows the evolution of the principal parameters of V838~Mon
with time. The effective temperature, luminosity and effective radius from
Table~\ref{evol_t} are displayed in the upper, middle and bottom panel, respectively.
On the abscissa in each panel the epoch of observations (days since
1~Jan.~2002), in the logarithmic scale is given. Star-like symbols  
show the results obtained from fitting the supergiant spectra to the $BVRIJHKL$ 
(or $BVRIJHK$ for epochs $> 300$~days) magnitudes,
i.e. those labelled with $BL$ or $BK$ in column (2) of Table~\ref{evol_t}.
The same but derived from the $BVRI$ magnitudes (labelled $BI$ in
column (2) of Table~\ref{evol_t}) is represented with full dots.
Open symbols display the results obtained from fitting the blackbody spectra, i.e.
those labelled BB in column (3) of Table~\ref{evol_t}. Dotted and dashed
curves are discussed Sect.~\ref{erup} and \ref{decline}.

As can be seen from Fig.~\ref{evol_f}, in {\it pre-eruption} and {\it eruption}
(log~time~$\la 2.0$), there is no significant systematic 
difference between the results obtained from fitting only the
$BVRI$ magnitudes and those
derived form the $BVRIJHKL$ measurements. The full dots 
consistently follow the same evolutionary trend as the asterisks.
In the {\it decline} phase, given the uncertainties due to the reference spectra,
the results obtained from fitting the blackbodies (open circles) are fairly 
consistent with those based on the supergiant spectra (asterisks).

In various estimates done later in this paper, requiring parameters of
the central star, we adopt, following Tylenda et~al. (\cite{tss}), standard
values for a B3\,V star, i.e. $M_\ast = 8~M_{\sun}$ and $R_\ast = 5~R_{\sun}$.

\subsection{{\it Pre-eruption}: epochs earlier than 32 days  \label{pre-erup}}

In the {\it pre-eruption} phase (log~time~$\la$~1.5 in Fig.~\ref{evol_f}) 
V838~Mon seemed to have 
reached an equilibrium or relaxation configuration.
This phase followed an initial
event which had presumably taken place a few days before 1~January~2002
(e.g. Munari et~al. \cite{munari}).
The photospheric radius between days 10--32 remained
at an almost constant value of $\sim 350~R_{\sun}$, suggesting that
the photosphere was in a quasi-hydrostatic equilibrium. The effective temperature
reached a maximum at days 12--15 and subsequently slowly declined.
The luminosity showed a similar trend and at the end of the {\it pre-eruption} 
phase the objects was half as luminous as at the beginning. 

This behaviour
suggests that the initial event, which inflated the stellar envelope to
$\sim 350~R_{\sun}$, had properties of a short-lived energy burst, after which
the energy sources dropped significantly, provoking a cooling of the photospheric
regions and a decline in the apparent luminosity.
In principle, cooling and a luminosity decline
should be followed by a contraction of the photosphere.
However, for an $8~M_{\sun}$ star
and the observed photospheric radius the dynamical time is $\sim 30$~days.
Thus the inflated stellar envelope did not have enough time to react during
the {\it pre-eruption} phase.

Within the above hypothesis one can estimate the mass of the inflated envelope
assuming that during the {\it pre-eruption} phase it radiated away a significant
part of its internal energy. 
The latter can be estimated from Eq.~(\ref{e_3/2}).
Integrating the observed 
luminosity over the {\it pre-eruption} phase and assuming (unperturbed)
stellar parameters, $M_\ast = 8~M_{\sun}$ and $R_\ast = 5~R_{\sun}$, one gets 
a mass of the inflated envelope, $M_\mathrm{e} \simeq 4 \times 10^{-3} M_{\sun}$.
This value should be considered with caution. On the one hand, the observed luminosity and
temperature evolved only slightly during {\it pre-eruption}, so
only a part of the envelope energy radiated away.
On the other hand, it is likely the energy sources that had presumably initiated 
{\it pre-eruption} did not extinguish completely and were contributing
to the observed luminosity in the course of the {\it pre-eruption} phase.

Spectroscopic observations obtained
in the {\it pre-eruption} phase showed numerous lines with P-Cygni profiles
indicating significant mass loss.
The typical outflow velocity was of 200--300~$\mbox{km}\ \mbox{s}^{-1}$ with 
a terminal one reaching
500~$\mbox{km}\ \mbox{s}^{-1}$ (Munari et~al. \cite{munari}, 
Osiwa\l{}a et~al. \cite{osiwal}).
This mass loss, observed during the whole {\it pre-eruption} phase, is not surprising,
as the stellar luminosity was then of a quarter of the Eddington limit for 
an 8~$M_{\sun}$ star.

\subsection{{\it Eruption}: epochs between 33 and $\sim 100$ days  \label{erup}}

The gentle relaxation in the {\it pre-eruption} phase was suddenly interupted 
on 2~February~2002 (day 33). The event was very dramatic. 
For almost 3 days V838~Mon
brightened by a factor of $\sim 25$ in luminosity while the effective temperature
increased from $\sim 4600$~K to $\sim 7200$~K. The event was also followed by
a fast rise of the effective radius. A major qualitative difference between
the evolution of V838~Mon in the whole {\it eruption} phase compared to that in 
{\it pre-eruption} is a more or less steady increase of the effective radius
in {\it eruption} versus a roughly constant radius in {\it pre-eruption}. At the
end of the {\it eruption} phase the effective radius reached values approaching
$3000~R_{\sun}$, compared to $\sim 350~R_{\sun}$ in {\it pre-eruption}.

However, as can be seen from Fig.~\ref{evol_f} (bottom panel), the rise of the
effective radius was not monotonic. One can easily distinguish
three periods during which the radius was steadily increasing, interupted by two
short phases of contraction. A simplistic interpretation of this behaviour might be
that the effective radius evolution reflects the behaviour of the stellar envelope,
namely that each phase of the envelope expansion is followed by a period during 
which the envelope is contracting. However, the expansion velocities as 
estimated below are much higher than the local escape velocities. Thus there is no way
to stop the expanding matter at a certain time and to force it to contract. 
A more likely hypothesis is that during the contraction phase the hitherto 
photospheric layers become, due to expansion, optically thin and the photosphere 
starts to form in deeper denser regions. In this case the observed 
evolution of the effective radius during {\it eruption} would suggest ejection of 
three consecutive shells.

To futher investigate the evolution of the effective radius  
we have plotted three dotted lines
in the bottom panel of Fig.~\ref{evol_f} showing radii increasing with constant
velocity. The parameters of the lines, i.e. velocity and epoch of the zero
radius, are 800~$\mbox{km}\ \mbox{s}^{-1}$ and 28.4~days, 
400~$\mbox{km}\ \mbox{s}^{-1}$ and 30.5~days, 270~$\mbox{km}\ \mbox{s}^{-1}$ and 28.0~days
for lines ({\bf a}), ({\bf b}) and ({\bf c}), respectively. The fact that the three
lines start at almost (within 2.5 days) the same time moment strongly suggests that
the whole {\it eruption} phase resulted from a single outburst event which took
place in the last days of January~2002, deep in the stellar envelope already inflated
by the {\it pre-outburst} event.

The above suggestion that during the {\it eruption} phase V838~Mon ejected three
consecutive shells poses some problems when spectroscopic data are considered. 
The velocity of the first shell (line ({\bf a}) in the bottom panel of 
Fig.~\ref{evol_f}) of 800~$\mbox{km}\ \mbox{s}^{-1}$ is much higher than the observed values. In general,
the observed velocities during {\it eruption} were similar or even lower 
than during {\it pre-eruption}. 
The terminal velocity derived by Kipper et~al. (\cite{kipper}) on 4/5~Feb.~2002
(days 35--36) was  between 240 and 390~$\mbox{km}\ \mbox{s}^{-1}$. 
Even if the systematic velocity
of the object, $\sim 65\ \mbox{km}\ \mbox{s}^{-1}$ as discussed in Sect.~\ref{ext_d}, 
is taken into account, this is
much lower than the velocity of our hypothetical first shell.
Typical velocities of maximum absorption in the P-Cygni profiles in {\it eruption} 
were between 60 and 300~$\mbox{km}\ \mbox{s}^{-1}$ with a general tendecy of decreasing
(or the appearance of lower velocity components) with time
(Kolev et~al. \cite{kolev}, Crause et~al. \cite{lisa03},
Wisniewski et~al. \cite{wisnia}, Kipper et~al. \cite{kipper}). If the systematic
velocity of the object is added, these values are consistent with the velocities
of the shells ({\bf b}) and ({\bf c}) in Fig.~\ref{evol_f}.

A varying photospheric radius need not reflect a mouvement of the matter
in the photospheric layers. As discussed above, the shrinkage of the photosphere,
observed on days $\sim 38$ and $\sim 60$, cannot be considered as due to a collapse of
the envelope. Similarily an increase of the photospheric radius does not
necessarily require expansion of a dense, optically thick shell.
In the case of an optically thick wind, e.g. as discussed in
Bath \& Shaviv (\cite{bs76}) for classical novae, the photosphere is formed in
the wind and its position is mainly determined by the mass loss rate, $\dot{M}$.
If $\dot{M}$ is increasing the photosphere moves outward, while a decrease
in $\dot{M}$ leads to photospheric contraction. However, even if $\dot{M}$ remains
constant, but the temperature of the wind matter varies, the position
of the photosphere would also vary because of the varying opacity.
We argue that this is what happened in the luminosity peak 
observed during the first $\sim 10$~days of the {\it eruption} phase.

Let us assume a steady-state spherically symmetric wind with a mass loss rate,
$\dot{M}$, and expanding with a velocity, $v_\mathrm{w}$. Then the density, $\rho$, 
of the wind matter is distributed with a radius, $r$, as
\begin{equation}
  \rho = \frac{\dot{M}}{4 \pi\,r^2\,v_\mathrm{w}}.
\label{ro_eq}
\end{equation}
The effective photosphere is formed at a radius, $R_0$, satisfying the condition
\begin{equation}
  \int_{R_0}^{\infty} \kappa\,\rho\,dr = \frac{2}{3},
\label{r0_eq}
\end{equation}
where $\kappa$ is the opacity. Assuming a constant mass loss rate, velocity 
and opacity above the photosphere, Eq.~(\ref{r0_eq}) can be integrated,
using Eq.~(\ref{ro_eq}), leading to
\begin{equation}
  \kappa\,\rho_0\,R_0 = \frac{2}{3},
\label{phot_eq}
\end{equation}
where $\rho_0 = \dot{M}/(4\pi\,R_0^2\,v_\mathrm{w})$ is the density in the
photospheric region. For a given chemical composition, the opacity, $\kappa$, 
is a function of temperature and density. 
Therefore, if values for the effective temperature
and radius are given, the density, $\rho_0$, can be derived from Eq.~(\ref{phot_eq}).
Next, using the wind velocity, $v_\mathrm{w}$, the mass loss rate
$\dot{M}$ can be calculated from Eq.~(\ref{ro_eq}).

\begin{figure}
\centering
  \resizebox{\hsize}{!}{\includegraphics{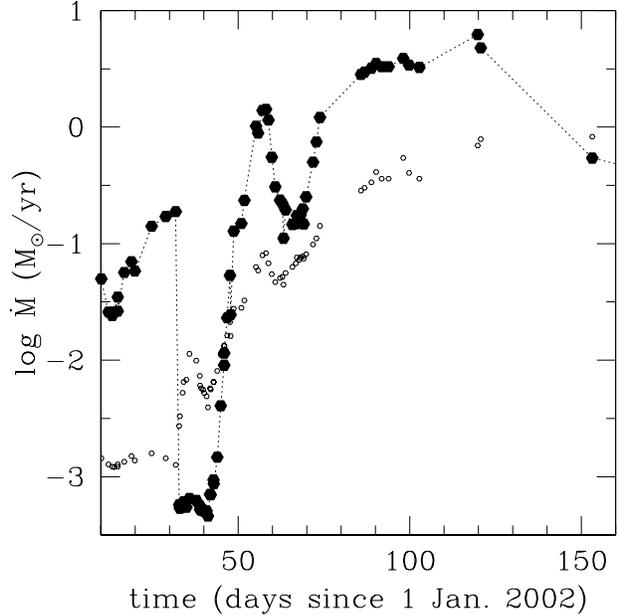}}
  \caption{Mass loss rate derived assuming an optically thick wind model
(see text for more details) -- large symbols. Small symbols -- effective
radius (the same as in Fig.~\ref{evol_f} but plotted as 
3~log$(R/R_{\sun}) - 10.5$).} 
\label{m_loss_f}
\end{figure}

Figure \ref{m_loss_f} shows the mass loss rate derived from the values of 
the effective temperature and radius listed in Table~\ref{evol_t}, using
the above steady-state wind model.
As the opacity we have used the Rosseland means
as given by Alexander \& Ferguson (\cite{af94}) for the chemical composition
$X =0.7$ and $Z = 0.02$.
A constant wind velocity, $v_\mathrm{w} = 300$~$\mbox{km}\ \mbox{s}^{-1}$, has been assumed.
To make the discussion easier the effective radius has also been plotted
in the figure with small symbols.

The values of $\dot{M}$ plotted in Fig.~\ref{m_loss_f}
are reasonable estimates of the mass loss rate only if the wind is optically thick
(so the effective photosphere is formed in the wind) and builds up an $\sim r^{-2}$ 
density distribution above the photosphere. The latter condition takes place
if the wind does not vary significantly on a time scale $\sim R_0/v_\mathrm{w}$.
This time scale is $\sim 10$~days in {\it pre-eruption}. Therefore the very
fast drop in mass loss rate, by more than 2 orders of magnitude on a time scale
of $\la$~1~day, obtained on day 32 is not real. If the wind
in {\it pre-eruption} was optically thick and the drop in mass loss rate 
was real, the density distribution above the photosphere would have significantly
changed after $\sim 10$~days. However, this was the beginning of
the {\it eruption} phase and the rise in luminosity should be followed
by an increase in mass loss rate rather than a drop. However what is striking
in Fig.~\ref{m_loss_f} is a constant mass loss rate obtained during the first
luminosity peak in {\it eruption}, i.e. between days 32 and 42. During
these 10 days all the stellar parameters, i.e. luminosity, effective temperature
and radius, varied by significant factors. For instance, the luminosity
first increased by a factor of 25 and then dropped by a factor of 2. Yet the mass loss
rate was constant within 10--15\%. We therefore suggest that this was
the time period during which the wind was optically thick and our model
of a steady-state wind gives a good interpretation of the evolution 
of the object. However, the
infered mass loss rate charactarises the wind during the {\it pre-eruption}
phase rather than in {\it eruption}. Thus we interepret the results
obtained in {\it pre-eruption} and {\it eruption} as follows.

In the {\it pre-eruption} phase, as discussed in Sect.~\ref{pre-erup}, the photosphere
was formed in layers approximately in hydrostatic equilibrium. The object had
a wind as indicated by the observed P-Cygni profiles. However the mass loss rate
was much lower than that shown in Fig.~\ref{m_loss_f} during this phase.
Instead we argue that $\dot{M}$ was then close to the value derived from the
analysis of the first 10~days in {\it eruption}, i.e. it was
$\sim 5 \times 10^{-4} M_{\sun}$/yr. Note that, as discussed in 
Sect.~\ref{pre-erup}, the luminosity of the object in {\it pre-eruption} 
was of a quarter of the
Eddington limit so a significant mass loss could be expected.
At the above mass loss rate and a photospheric temperature
of $\sim 5000$~K, the wind, during the {\it pre-eruption} phase,
was optically thin and seen only in the lines.
This roughly steady-state wind built up an $\sim r^{-2}$ density distribution
above the photosphere. On day 32 a strong luminosity wave
(or a series of strong shock waves),
presumably generated at the base of the inflated envelope, reached the photosphere
and quickly heated up the regions above it. This resulted in an abrupt increase of 
the opacity (by 2 orders of magnitude if the temperature increased from
5000~K to 7000~K -- Alexander \& Ferguson \cite{af94}) and the hitherto optically 
thin regions built by the wind became optically thick. This was observed as
a very fast expansion of the photospheric radius leading to an apparent
velocity of $\sim 800\ \mbox{km}\ \mbox{s}^{-1}$, 
much faster than the wind velocity observed in 
the lines. Subsequent cooling of the wind matter decreased the opacity, 
which resulted in a contraction of the effective photosphere observed between 
days 36 and 42. 

The luminosity wave was followed by an intense mass loss (the observed
luminosity between days 33 and $\sim 110$ was above the Eddington limit
for an $8~M_{\sun}$ star) which reached the shrinking photosphere near day 42,
as indicated by a large increase of $\dot{M}$ in Fig.~\ref{m_loss_f}. Note, however,
that the obtained values of $\dot{M}$ should again be regarded with caution. 
The assumption of an $r^{-2}$ density distribution in the model wind is not
satisfied if the mass loss rate is rapidly changing, as seen in Fig.~\ref{m_loss_f}
after day 42. In this case the obtained evolution of $\dot{M}$ in
the later epochs, when combined with the evolution of the effective radius, seems
to suggests that mass loss following the fast luminosity wave
took place in two shells. These shells were presumably
formed by the initial eruption event which took place at the end
of January, as discussed above, and followed the initial strong 
luminosity wave.
Interaction of the luminosity wave with the
envelope inflated in {\it pre-eruption} also might have been important for
the formation of the shells. 

The first shell, somewhat faster than the second one, 
became visible near day 42, when it met the photosphere contracting
in the {\it pre-eruption} wind.
Subsequently, it was observed as an expanding 
and cooling photosphere radiating at a roughly constant luminosity of 
$\sim 6 \times 10^5~L_{\sun}$. Due to expansion and cooling this shell became 
partly transparent around day 58 and 
the second shell started to be visible. This resulted in a shrinkage of the
effective photosphere observed between days 58 and 63. At the same time
the photosphere became hotter while the luminosity started to increase and
quickly reached a value of $\sim 1 \times 10^6~L_{\sun}$
remaining more or less constant between days
66 and 100. Judging from the evolution of the effective radius and luminosity,
the second shell became partly transparent between days 100--120. The discrepancy in
the effective radius derived from fitting supergiant spectra and blackbodies
on day 120 (see Table~\ref{evol_t} and Fig.~\ref{evol_f}) can be explained
by the presence of an extended, partly transparent envelope. At a temperature
of 2500--3000~K the matter is opaque in the optical due to molecular
absorption so the effective radius derived from the $BVR$ bands (important when
fitting a supergiant spectrum) would correspond to the outer radius of the envelope.
In the infrared the envelope is more transparent and it is possible to see
deeper and hotter layers. Thus a higher effective temperature and 
a smaller effective radius is obtained from fitting
a blackbody to the $IJHKL$ measurements.

From the parameters observed when a shell becomes transparent one can estimate
the mass of the shell, as the optical thickness of the shell is then expected
to be
\begin{equation}
  \tau = \kappa\,\rho\,\Delta R_\mathrm{s} \simeq \frac{2}{3}
\label{shell_eq}
\end{equation}
where $\Delta R_\mathrm{s}$ is the radial thickness of the shell. 
If $\Delta R_\mathrm{s}$, 
temperature, $T$, and $\kappa(T,\rho)$ are known, the density, $\rho$, can be 
derived from Eq.~(\ref{shell_eq}). Then, knowing the shell radius, 
$R_\mathrm{s}$, the mass of the shell, $M_\mathrm{s}$, can be obtained.

As discussed above the first shell started to become partely transparent on
day $\sim 58$. We can thus assume the effective radius and temperature obtained
at this epoch (see Table~\ref{evol_t}) as estimates of the shell radius and 
temperature, i.e.
$R_\mathrm{s} = 1.3 \times 10^3~R_{\sun}$ and $T = 4500$~K. From a difference between
curves ({\bf b}) and ({\bf c}) in the bottom panel of Fig.~\ref{evol_f} at this epoch,
we can estimate that $\Delta R_\mathrm{s} \simeq 0.2~R_\mathrm{s}$. 
Adopting the Rosseland
mean opacity as tabulated in Alexander \& Ferguson (\cite{af94}) we finally 
obtained $M_\mathrm{s} \simeq 0.08~M_{\sun}$.

The second shell presumably became transparent between days 100--120. 
Assuming that Eq.~(\ref{shell_eq}) was satisfied on day 100 we have 
(see Table~\ref{evol_t}) $R_\mathrm{s} = 2.5 \times 10^3~R_{\sun}$ and $T = 3500$~K,
which results in $M_\mathrm{s} \simeq 0.45~M_{\sun}$. If we assume that the shell
became partly transparent on day 120 and that the results obtained from fitting
a supergiant spectrum (see Table~\ref{evol_t}) are estimates of the shell parameters, 
i.e. $R_\mathrm{s} = 3.0 \times 10^3~R_{\sun}$ and $T = 2500$~K, the result is
$M_\mathrm{s} \simeq 0.65~M_{\sun}$.

The above estimates show that the second shell was $\sim 7$ times more massive than
the first one. However, the obtained absolute values
probably overestimate the real shell masses.
The reason is that the Rosseland mean, which very well approximates the true
opacity in optically thick conditions, is likely to underestimate the real opacity
in partly transparent media, particularly in conditions where
the Planck mean is much higher than the Rosseland mean. This is the case for the
temperature and density range considered above where, according to the results of
Alexander \& Ferguson (\cite{af94}), the Planck mean is $\sim 2$ orders of magnitude 
higher than the Rosseland mean. If the Planck mean is used instead of the Rosseland
mean in Eq.~(\ref{shell_eq}) the results are
$M_\mathrm{s} \simeq 8 \times 10^{-4}~M_{\sun}$ and $\simeq 4 \times 10^{-3}~M_{\sun}$
for the masses of the first and second shell, respectively. These values can be considered
as lower limits to the shell masses.

\subsection{{\it Decline}: epochs later than $\sim 100$ days  \label{decline}}

This phase is first marked by a continuous decline in luminosity.
This decline was initially fast. For about 50~days V838~Mon faded by
an order of magnitude. Next the time scale became longer
and after almost three years the object was $\sim 40$ times less luminous than
in {\it eruption}. 

The initial {\it decline}, untill day $\sim 230$, was also characterised by 
a decreasing effective temperature which dropped to $\la 2000$~K. The results
of our analysis during this phase are particularly uncertain. The spectral fitting
was based on an uncertain extrapolation well beyond the range of the standard supergiant
spectra. Also the blackbody spectra were not successful in fitting the optical
and infrared magnitudes at the same time. The evolution of the effective radius
was particularly uncertain, as can be seen from the bottom panel of 
Fig.~\ref{evol_f}.

In the later {\it decline}, i.e. days $\ga 250$, the object shrunk
in effective radius. Also the trend in the effective temperature
was reversed. The object, although still very cool for a star, 
slowly became hotter. If the observed evolution of
V838~Mon is plotted on the HR diagram the conclusion is clear:
in the later {\it decline} the object declined along the Hayashi track.
The only known stellar eruptive object that shows a similar evolution
is V4332~Sgr (Tylenda et~al. \cite{tcgs}) which is
usually considered to be of the same class as V838~Mon.

In the stellar evolution the only case of decline along the Hayashi track
is the protostellar phase when the main source of energy is gravitational
contraction. This strongly suggests that V838~Mon in the late {\it decline}
also is powered by a gravitational collapse of its envelope inflated during
{\it eruption}. As shown in Tylenda et~al. (\cite{tcgs}) this hypothesis can
satisfactorily explain the observed evolution of V4332~Sgr for almost 10~years
after its eruption. Within this hypothesis one can estimate the mass of 
the inflated envelope. 

Appendix \ref{contrac} presents an approach to the problem of a gravitationally 
contracting envelope. The structure of the envelope is treated in a polytropic 
approximation which seems to be better than the $r^{-5/2}$ density distribution
assumed in Tylenda et~al. (\cite{tcgs}).
For the effective temperatures observed during {\it decline} it is
reasonable to assume that the envelope is convective so $n = 3/2$.
The dashed curves in the middle and bottom panels of Fig.~\ref{evol_f} 
were derived from a numerical integration of Eq.~(\ref{dr_env})
(using Eqs.~\ref{im_3/2} and \ref{ie_3/2}).
We found that $T_\mathrm{eff} = 6120 - 1200\,\log (R/R_{\sun})$
is a good approximation to our results obtained from fitting the supergiant spectra
to the observations for days $\ga 250$. This relation was used in our modelling
of the envelope contraction.
$M_\ast = 8~M_{\sun}$ and $R_\ast = 5~R_{\sun}$ have been assumed
as the parameters of the central star.
The mass of the envelope was chosen to fit
the observed radius and in the case of the dashed curves in Fig.~\ref{evol_f}
it is $M_\mathrm{e} = 0.22~M_{\sun}$.

\section{Infrared excess  \label{ire}}

Infrared spectra of V838~Mon have been
analysed by Lynch et~al. (\cite{lynch04}). 
The observations taken in January~2002 (i.e. in {\it pre-eruption})
they have been able to fit with a single
blackbody of 2700~K. They, however, note a possible small excess near 
10~$\mu$m. The data obtained a year later, between December~2002 and 
February~2003 (i.e. in {\it decline}), show a more complex nature. 
In particular, the spectrum shows strong absorption
bands due to numerous molecules characteristic of an oxygen-rich (C/O~$<$~1) matter.
Lynch et~al. (\cite{lynch04}) conclude that the overall spectrum 
at short wavelengths can then be fitted with a 2000~K blackbody,
whereas at longer wavelengths (8--13~$\mu$m) the continuum indicates 650--800~K.

\begin{figure*}
  \centering
  \includegraphics[width=8.9cm]{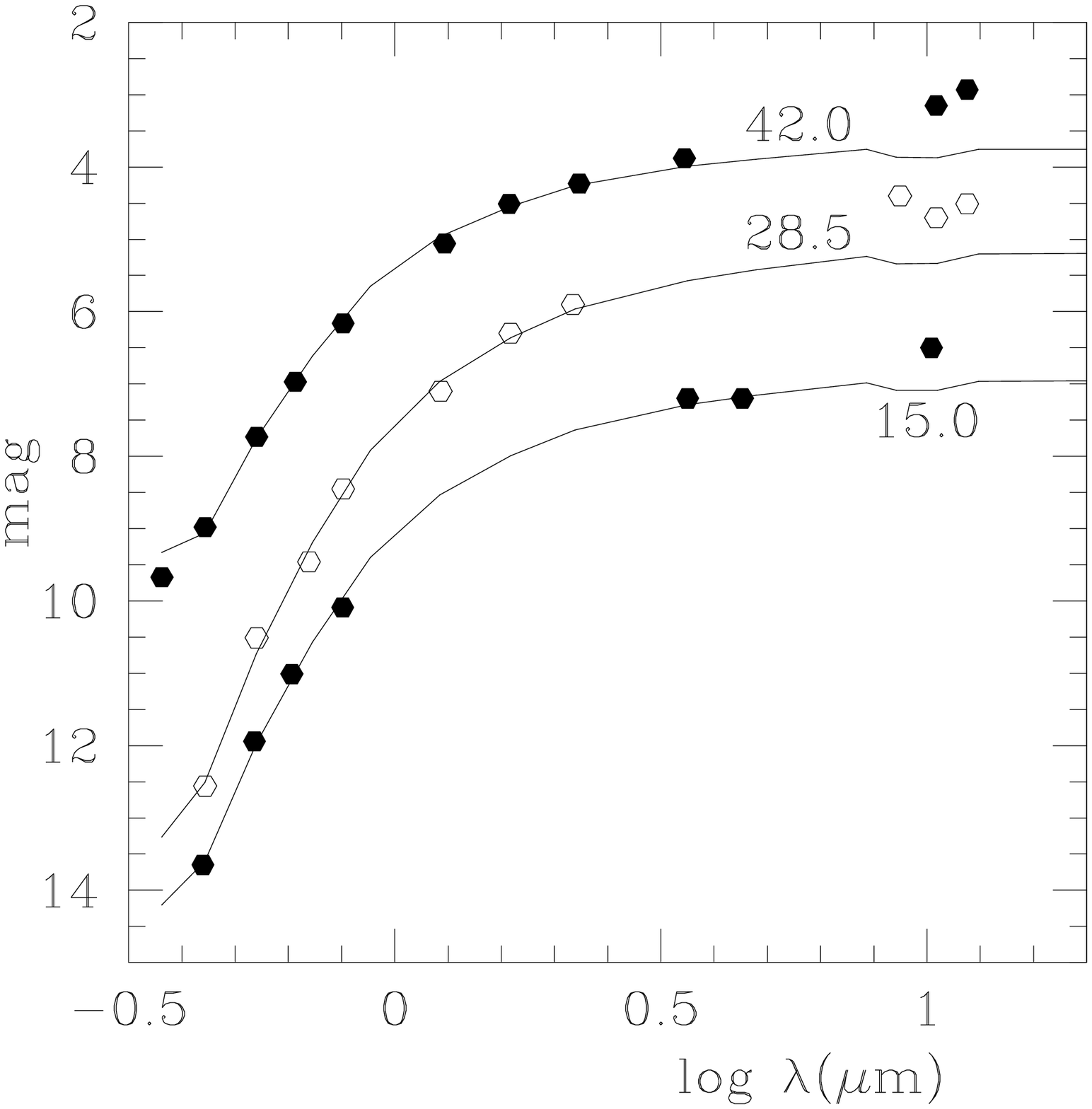}
  \includegraphics[width=8.9cm]{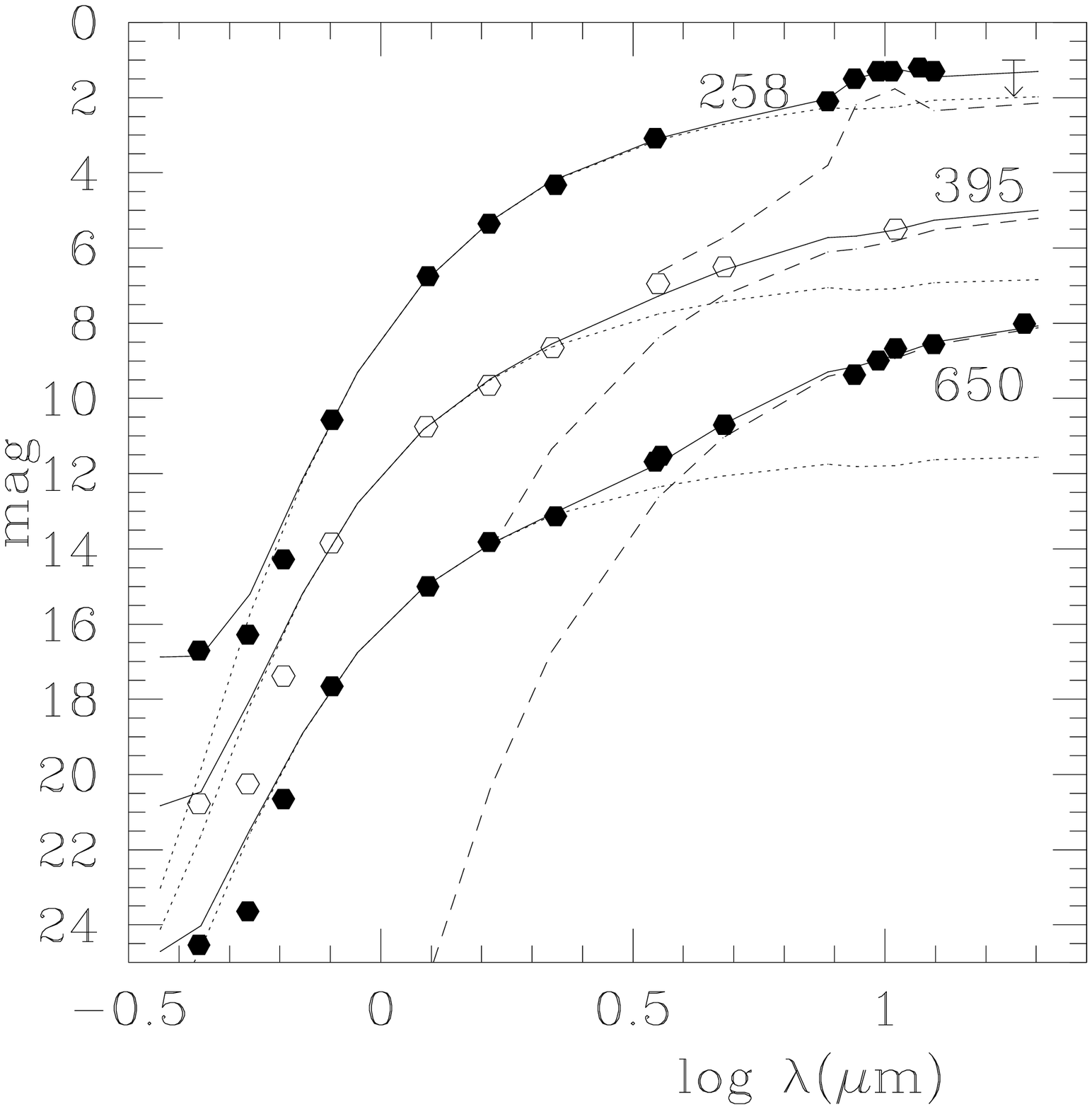}
  \caption{Infrared excess in {\it pre-eruption} and {\it eruption} 
({\bf left panel}) as well as in {\it decline} ({\bf right panel}). Symbols -- 
observed magnitudes. Full curves -- fitted spectra.
Dotted and dashed curves (right panel) -- stellar and dust spectral components, 
respectively.
The curves are labelled with
the epoch of observations (days since 1~Jan.~2002). A downward arrow for day 258 
shows the upper observational limit at 18~$\mu$m.
Note that the curve and the data
on day 15.0 (left panel) have been shifted by 2 magnitudes,
while those  corresponding to days 395 and 650 (right pannel) 
have been shifted by 4 and 8 magnitudes, respectively.
The data sources and the parameters of the fitted spectra can be found in
Table~\ref{ir_t}.}
  \label{fig_ir}
\end{figure*}

We have collected the available measurements in the range of $\lambda \ga 5~\mu$m.
With the shorter wavelength photometry (discussed in Sect.~\ref{results}) this allowed
us to investigate the presence of an infrared excess for six epochs, as given in
column (1) of Table~\ref{ir_t}. Two of them are in
{\it pre-eruption}, one in {\it eruption} and three in {\it decline}. The data and 
results are summarized in Table~\ref{ir_t} and Fig.~\ref{fig_ir}.

\begin{table}
\centering
\caption{Infrared excess in V838~Mon: data and results.
See text for explanations of the columns.}
\label{ir_t}
\begin{tabular}{@{} p{0.4cm} p{2.15cm} p{0.6cm} p{0.6cm} p{0.6cm} p{0.8cm} p{0.7cm} @{}}
\hline
  day & data & $T_\mathrm{BB,s}$ & $T_\mathrm{BB,d}$ & $-$lg\,$\theta_\mathrm{d}$ & 
$R_\mathrm{d}/R_{\sun}$ & $L_\mathrm{d}/L_\mathrm{s}$ \\
\hline
   ~15. & $BVRI$(clk3) & 5200. \\
        & $LMN$(lrr4) \\
   28.5 & $BVRI$(kls2) & 4650. \\
        & $JHK$(lrr4) \\
        & 9--12$\mu$m(klk2) \\
   ~42. & $U$--$L$(mhk2) & 6880. \\
        & 10--12$\mu$m(klk2) \\
  258. & $B$--$L$(clm5) & 1840. \\
        & 8--18$\mu$m(wc2) \\
  395. & $B$--$L$(clm5)  & 2220. &  800. & ~~7.75 & ~~6.3E3 & 0.29 \\
        & $J$--$N$(lrr4) \\
  650. & $B$--$L$(clm5)  & 2520. &  600. & ~~7.43 & 13.1E3 & 0.54 \\
        & $LMN$(lrp3) \\
        & 9--19$\mu$m(tp3) \\
\hline
\end{tabular}
\\
\begin{flushleft}
Data references: 
clk3 -- Crause et~al. (\cite{lisa03}),
clm5 -- Crause et~al. (\cite{lisa04}),
klk2 -- Kaeufl et~al. (\cite{kaufl}),
kls2 -- Kimeswenger et~al. (\cite{kimes}),
lrp3 -- Lynch et~al. (\cite{lynch03}),
lrr4 -- Lynch et~al. (\cite{lynch04}),
mhk2 -- Munari et~al. (\cite{munari}),
tp3 -- Tapia \& Persi (\cite{tapia}),
wc2 -- Watson \& Costero (\cite{watcos})
\end{flushleft}
\end{table}

Standard supergiant colours are poorly known for the $MN$ bands. Therefore
in order to determine the stellar contribution for $\lambda \ga 5~\mu$m
we have used blackbody distributions fitted to shorter wavelengths. For the
{\it decline} period these are simply the same blackbody spectra as those discussed
in Sect.~\ref{results} and whose parameters are given in Table~\ref{evol_t}. 
They are shown with dotted curves 
in the right panel of Fig.~\ref{fig_ir} .
In the case
of {\it pre-eruption} and {\it eruption} we have fitted blackbodies to the
observed magnitudes shortward of 5~$\mu$m. They are shown with full curves in
the left panel of Fig.~\ref{fig_ir}. The values of the stellar blackbody
temperatures, $T_\mathrm{BB,s}$, are given in column (3) of Table~\ref{ir_t}. 
Note that the blackbody
temperatures obtained for the first three epochs in Table~\ref{ir_t} are not
significantly different from the values of $T_\mathrm{eff}$ obtained from fitting
supergiant spectra (see Table~\ref{evol_t}).

As can be seen from Fig.~\ref{fig_ir} (left panel) the data in {\it pre-eruption}
and {\it eruption} show a 0.5--1.0 magnitude excess at $\sim 10~\mu$m. This is
in agreement with the analysis of Lynch et~al. (\cite{lynch04}) of their
January~2002 spectrum.\footnote{The difference in the blackbody temperature
between Lynch et~al. (\cite{lynch04}) and the present study 
(2700~K versus $\sim 5000$~K) is simply due to the interstellar reddenning correction, 
i.e. no reddenning correction
in Lynch et~al, versus $E_{B-V} = 0.9$ in the present study. Indeed, if we apply
our fitting procedure with $E_{B-V} = 0.0$ to the $JHKLM$ magnitudes of Lynch et~al.
in January~2002, the resultant temperature is $\sim 2700$~K.} As noted in Lynch et~al. 
the 10~$\mu$m excess is likely to be due to silicate emission.

A similar conclusion can also be drawn from a {\it decline} spectrum
of day 258 (mid-September~2002). For this epoch we have more detailed measurements
near 10~$\mu$m from Watson \& Costero (\cite{watcos}).
As can be seen from Fig.~\ref{fig_ir} 
(right panel), the stellar blackbody fits the data well up to 
$\sim 8~\mu$m. Above this wavelength the observed flux steeply increases and between 
9--13~$\mu$m it stays $\sim 0.8$~magnitude above the stellar continuum. This
feature can indeed be interpreted as due to optically thin emission from 
silicate dust. To show this we have added a component proportional to the dust
emission coefficient, i.e. to $Q_{\mathrm{abs},\nu}\,B_\nu(T_\mathrm{d}$), where
$Q_{\mathrm{abs},\nu}$ is a monochromatic absorption coefficient for
"astronomical silicates" as tabulated in Draine (\cite{draine}). This component
is shown as a dashed curve on day 258 in Fig.~\ref{fig_ir}.
The full curve presents the sum of the stellar blackbody and the dust contribution
and, as can be seen from the figure, it fits the observations very well.
The dust temperature, $T_\mathrm{d}$, has been assumed to be 800~K as this is
the expected temperature of silicate dust condensation. This value is also close to 
the blackbody temperature estimated in Lynch et~al. (\cite{lynch04}) and in the
present study from the observations in December~2002 -- January~2003 (day 395,
see below). The value of $T_\mathrm{d}$ cannot be precisely constrained from
the observational data on day 258, however it cannot be lower than 
$\sim 400$~K as then the predicted brightness at 18~$\mu$m is above the observational
upper limit (shown by a downward arrow in Fig.~\ref{fig_ir}).

We can therefore conclude that in {\it pre-eruption}, {\it eruption} and early
{\it decline} V838~Mon showed an infrared excess at $\sim 10~\mu$m. This excess
was not significant for the global energetics of the object and was likely to be 
due to silicate dust condensation in the wind. The dust was probably optically thin,
suggesting a moderate mass loss rate (at the distance of dust condensation).
Most likely, following the discussion in Sect.~\ref{erup}, this concerns
the {\it pre-eruption} wind matter expanding above the two dense shells ejected
at the onset of {\it eruption}.

In later epochs the situation changed considerably. On days 395 (data from observations
in December~2002 and January~2003) and 650 (October~2003) 
the infrared excess increased
and cannot be fitted by an optically thin silicate emission.
In the former epoch
Lynch et~al (\cite{lynch04}) note a silicate feature at $\sim 10~\mu$m but 
it is superimposed on a blackbody of 650--800~K. In the latter epoch,
as noted in Lynch et~al. (\cite{lynch03}), the feature became even less pronounced
with a continuum emission dominating the observed spectrum.

We have been able to fit the whole spectrum of V838~Mon on days 395 and 650
by a sum of two blackbody distributions, as shown in Fig.~\ref{fig_ir} (right panel).
The parameters of the fits can be found in Table~\ref{ir_t}. The cooler component
(dashed curves in the figure)
can be interpreted as due to an optically thick dusty region reprocessing the
radiation from the central object. The fact that we see the central object
signifies that dust does not significantly obscure it. In other words,
the covering factor of the dusty region is $< 1$, in accordance with the discussion below.

As can be seen from Table~\ref{ir_t},
the dust temperature, $T_\mathrm{BB,d}$, decreased
while the radius, $R_\mathrm{d}$, increased 
between the two epochs. This can be easily interpreted as due to expansion of the
dusty region. 

The values of log~$\theta_\mathrm{d}$ and $R_\mathrm{d}$
given in Table~\ref{ir_t} are lower limits to the dust distance from the
central object, as they have been derived assuming a radiating sphere. 
The values of $L_\mathrm{d}/L_\mathrm{s}$ given in the last column
of Table~\ref{ir_t} can be considered as an estimate of the covering factor of
the dusty region. If this is the case, the dust distance should be increased by
the inverse square root of the covering factor and $R_\mathrm{d}$
becomes $1.17 \times 10^4~R_{\sun}$ and
$1.78 \times 10^4~R_{\sun}$ on days 395 and 650, respectively. 
These values of $R_\mathrm{d}$ and the derived temperature values 
closely follow the expected relation 
$T_\mathrm{BB,d}/T_\mathrm{BB,s} = (R_\mathrm{s}/R_\mathrm{d})^{1/2}$
($R_\mathrm{s}$ taken from Table~\ref{evol_t}). Assuming that, as discussed in
Sect.~\ref{erup}, the dominant mass loss was initiated on day $\sim 30$ one can 
estimate a mean expansion velocity from the above values of $R_\mathrm{d}$. 
The result is 260~$\mbox{km}\ \mbox{s}^{-1}$ and 230~$\mbox{km}\ \mbox{s}^{-1}$ 
for days 395 and 650, respectively.
We can therefore interpret the strong infrared excess observed in 
the later {\it decline} as due to dust formed in the second, more massive,
shell ejected in {\it eruption}, as discussed in Sect.~\ref{erup}.
From the condition that the shell should be optically thick 
we can estimate a lower limit to its mass. Following 
Alexander \& Ferguson (\cite{af94}) the opacity $\kappa \simeq 2.7$
for $T \simeq 700$~K (the Rosseland mean, but the Planck mean is not significantly different).
Taking the shell radius of $1.8 \times 10^4~R_{\sun}$, a radial thickness of
20\% and the filling factor of 0.5 the condition $\tau \ga 1$ implies
$M_\mathrm{s} \ga 2 \times 10^{-3}~M_{\sun}$. This value is consistent
with the estimates obtained in Sect.~\ref{erup} ($4 \times 10^{-3} - 0.6~M_{\sun}$).

The high value of
$T_\mathrm{BB,d}$ in Dec.~2002 -- Jan.~2003 (800~K) suggests that dust was
being formed during this epoch. Probably the whole process of
dust formation was completed in a somewhat later epoch. This might be the reason
why the covering factor as measured by $L_\mathrm{d}/L_\mathrm{s}$ on day 650
was larger than that on day 395.

\section{Summary  \label{sum}}

We present the following
description of the eruption of V838~Mon observed in 2002.

The eruption consisted of two distinct phases,
{\it pre-eruption} and {\it eruption}. Both phases are likely
to have resulted from rather short-lived but strong energy bursts. 

The first one, which produced the {\it pre-eruption}, occured
before the discovery of the object so we have no data for it.
Presumably it took place in the last days of December~2001
and its main effect was to produce an envelope inflated to $\sim 350~R_{\sun}$.
In January~2002 the object was in a relatively quiet phase, radiating at 
$\sim 7-8 \times 10^4~L_{\sun}$.
The slowly decreasing effective temperature and luminosity suggest
relaxation after the initial event. The object had a wind that built up an 
optically thin envelope, extending to $\sim 1500~R_{\sun}$ in last days
of January~2002.

Presumably, another eruption event occured on 28--30~January at the base of 
the inflated envelope. This second energy burst was much stronger than the first one
and generated the main phase, {\it eruption}.
As a result an energy wave, probably in the form of shocks and/or a radiation wave,
was formed, followed by a massive outflow of matter. The energy wave reached the
photosphere on 2~February and heated up the extended envelope built up by the
{\it pre-eruption} wind. This increased the opacity and made the inner wind layers optically
thick, which was observed as a rapid expansion of the effective photosphere.
In 3 days the effective radius increased to $\sim 700~R_{\sun}$, while the luminosity
reached $\sim 1 \times 10^6~L_{\sun}$. Subsequent cooling of the wind envelope 
resulted in a decrease in luminosity and a contraction of the photosphere. 
On 10--11~February the first shell of mass outflow met the contractiong photosphere
and became visible. This first, less massive shell, likely to have been produced from the
{\it pre-eruption} envelope compressed by the second energy burst, was observed as
an expanding and cooling photosphere untill $\sim 27$~February. Then it became partly 
transparent, so that deeper and hotter layers started to be visible. 
In this way the main mass
outflow was observed between 4~March and $\sim 10$~April
as an initially hotter, but then expanding and cooling, photosphere
radiating at a luminosity of $\sim 1 \times 10^6~L_{\sun}$. Thus we show that 
the evolution of the object in the {\it eruption} phase, i.e. in
February -- mid-April, was due to a single event rather than a series of outbursts as
often suggested in the literature.

Because of no significant
energy supply from internal sources the object declined in luminosity
when the shell of the 
main outflow became transparent and radiated away most of its internal energy.
A certain part of the matter involved in the eruption did not gain enough energy 
to overcome the gravitational field of the central star. It formed an inflated envelope
whose gravitational contraction was the main source of the luminosity observed in the
{\it decline} phase. We estimate that the mass of this envelope is $\sim 0.2~M_{\sun}$.

At the same time the matter ejected during {\it eruption} expanded and cooled,
so that dust formation was possible. As a result a strong infrared excess at 
$\lambda \ga 3~\mu$m was observed since October--November~2002. 
We estimate that the total mass lost from V838~Mon is $\la 0.6~M_{\sun}$.
This is rather a conservative upper limit. The true value is likely to be significantly
lower.

We postpone the discussion of the present results in terms of possible eruption
mechanisms to another paper.

\begin{acknowledgements}
The author thanks to Noam Soker for discussions and comments on the text of the paper.
This work was partly supported from grant No. 2.P03D.002.25 financed by the Polish State
Committee for Scientific Research.
\end{acknowledgements}

\appendix
\section{Polytropic envelope  \label{contrac}}

Suppose that a star of a mass, $M_\ast$, and a radius, $R_\ast$, has an
envelope of mass, $M_\mathrm{e}$, inflated to a radius, $R_\mathrm{e}$.
The envelope is in hydrostatic equilibrium.
Suppose also that $M_\mathrm{e} \ll M_\ast$, so the gravitational potential
within the envelope, $V(r)$, is
\begin{equation}
  V(r) = -\frac{G\,M_\ast}{r}.
\label{v}
\end{equation}

Let us assume a polytropic equation of state, i.e.
\begin{equation}
  P = K\,\rho^{1+(1/n)}, 
\label{politrop}
\end{equation}
where $P$ is pressure, $\rho$ -- density, $K$ -- polytropic constant, 
$n$ -- polytropic index. (A detailed theory of polytropes can be found
in e.g. Chandrasekhar \cite{chandra}.)
Then the standard equation of hydrostatic equilibrium can
be integrated giving
\begin{equation}
  (n+1)\frac{P}{\rho} = \frac{G\,M_\ast}{r} - \frac{G\,M_\ast}{R_\mathrm{e}}.
\label{P_ro}
\end{equation}
Eq. (\ref{P_ro}) can be used with Eq.~(\ref{politrop}) to obtain 
the density distribution, i.e.
\begin{equation}
  \rho = \rho_0 \left( \frac{x_0}{1-x_0} \right)^n \left( \frac{1-x}{x} \right)^n,
\label{ro}
\end{equation}
where $x = r/R_\mathrm{e}$, $x_0 = R_\ast/R_\mathrm{e}$, and $\rho_0 = \rho(x_0)$
is the density at the base of the envelope.
The mass of the envelope, $M_\mathrm{e}$, is
\begin{equation}
  M_\mathrm{e}\! =\! \int_{R_\ast}^{R_\mathrm{e}}\! 4\pi\,r^2\rho\,\mathrm{d} r
               = 4 \pi\,R_\mathrm{e}^3\,\rho_0 \left(\! \frac{x_0}{1-x_0} \!\right)^n
                 \!I_\mathrm{m}(n,x_0)
\label{envmass}
\end{equation}
where
\begin{equation}
  I_\mathrm{m}(n,x_0) = \int_{x_0}^{1} (1-x)^n\,x^{2-n}\,\mathrm{d} x .
\label{im}
\end{equation}
The potential energy of the envelope, $\Omega$, is
\begin{eqnarray}
\lefteqn{ \Omega = \int_{R_\ast}^{R_\mathrm{e}} 4\pi\,r^2\,V(r)\,\rho\, 
         \mathrm{d} r {} }
\nonumber\\
& & {}   = - 4 \pi\,G\,M_\ast\,R_\mathrm{e}^2\,\rho_0 
           \left( \frac{x_0}{1-x_0} \right)^n I_\mathrm{e}(n,x_0),
\label{omega1}
\end{eqnarray}
where
\begin{equation}
  I_\mathrm{e}(n,x_0) = \int_{x_0}^{1} (1-x)^n\,x^{1-n}\, \mathrm{d} x .
\label{ie}
\end{equation}
Using Eq.~(\ref{envmass}), Eq.~(\ref{omega1}) can be rewritten as
\begin{equation}
  \Omega = - \frac{G\,M_\ast\,M_\mathrm{e}}{R_\mathrm{e}} 
             \frac{I_\mathrm{e}(n,x_0)}{I_\mathrm{m}(n,x_0)}.
\label{omega2}
\end{equation}
According to the virial theorem the internal energy of the envelope, $U$, is
\begin{equation}
  U = - \frac{1}{2} \Omega.
\label{internal}
\end{equation}
Thus the total energy of the envelope, $E$, is
\begin{equation}
  E = \Omega + U = - \frac{G\,M_\ast\,M_\mathrm{e}}{2\,R_\mathrm{e}} 
             \frac{I_\mathrm{e}(n,x_0)}{I_\mathrm{m}(n,x_0)}.
\label{energy}
\end{equation}

In astrophysical applications two values of the polytropic index, i.e.
$n = 3/2$ and $n = 3$, are of particular interest. The integrals in
Eqs.~(\ref{im}) and (\ref{ie}) can be analytically evaluated in
these cases giving
\begin{eqnarray}
\lefteqn{ I_\mathrm{m}(3/2,x_0) = \frac{\pi}{16} - \frac{1}{8}\,\arcsin(x_0^{1/2})
         - \frac{1}{8}\,x_0^{1/2}\,(1-x_0)^{1/2} {} }
\nonumber\\
& & {} \qquad  - \frac{1}{12}\,x_0^{1/2}\,(1-x_0)^{3/2}
         + \frac{1}{3}\,x_0^{1/2}\,(1-x_0)^{5/2},
\label{im_3/2}
\end{eqnarray}
\begin{eqnarray}
\lefteqn{  I_\mathrm{e}(3/2,x_0) = \frac{3}{8}\pi - \frac{3}{4}\,\arcsin(x_0^{1/2})
         - \frac{3}{4}\,x_0^{1/2}\,(1-x_0)^{1/2} {} }
\nonumber\\
& & {} \qquad \qquad  - \frac{1}{2}\,x_0^{1/2}\,(1-x_0)^{3/2},
\label{ie_3/2}
\end{eqnarray}
\begin{equation}
  I_\mathrm{m}(3,x_0) = - \frac{11}{6} - \ln\,x_0 + 3 x_0 - \frac{3}{2} x_0^2
          + \frac{1}{3} x_0^3,
\label{im_3}
\end{equation}
\begin{equation}
  I_\mathrm{e}(3,x_0) = \frac{1}{x_0} + \frac{3}{2} + 3\,\ln\,x_0 - 3 x_0
          + \frac{1}{2} x_0^2.
\label{ie_3}
\end{equation}

Note that in the case of extended envelopes ($x_0 \ll 1$) the envelope energy
becomes 
\begin{equation}
  E \simeq - \frac{3\,G\,M_\ast\,M_\mathrm{e}}{R_\mathrm{e}} 
    \qquad  \mbox{for} \qquad n = 3/2
\label{e_3/2}
\end{equation}
and
\begin{equation}
  E \simeq - \frac{G\,M_\ast\,M_\mathrm{e}}{2\,R_\ast\,\ln(R_\mathrm{e}/R_\ast)} 
    \qquad \mbox{for} \qquad n = 3.
\label{e_3}
\end{equation}

Assuming that the luminosity of the envelope, $L$, is due to gravitational contraction
of the envelope and neglecting mass loss from the envelope, we can write
\newcommand{\ud}{\mathrm{d}}
\begin{eqnarray}
\lefteqn{ L = - \frac{\ud E}{\ud t} {} }\\
& & {}   =  \frac{G\,M_\ast\,M_\mathrm{e}}{2\,R_\mathrm{e}} 
      \left[ \frac{\ud}{\ud t} 
      \left(\! \frac{I_\mathrm{e}(n,x_0)}{I_\mathrm{m}(n,x_0)} \!\right)
     - \frac{I_\mathrm{e}(n,x_0)}{I_\mathrm{m}(n,x_0)}\,
       \frac{\ud}{\ud t}
       (\ln R_\mathrm{e}) \right].
\nonumber
\label{lum_g}
\end{eqnarray}
On the other hand, we have
\begin{equation}
  L = 4\pi\,R_\mathrm{e}^2\,\sigma\,T_\mathrm{eff}^4,
\label{lum_eff}
\end{equation}
where $T_\mathrm{eff}$ is the effective temperature.
Combining Eq. (\ref{lum_g}) and Eq.~(\ref{lum_eff}) we obtain
\begin{eqnarray}
\lefteqn{ \frac{\ud}{\ud t} (\ln R_\mathrm{e})
 = \frac{\ud}{\ud t} \left( \ln \frac{I_\mathrm{e}(n,x_0)}{I_\mathrm{m}(n,x_0)} \right)
{} }
\nonumber\\
& & {}  \qquad - \frac{8\pi\,R_\mathrm{e}^3\,\sigma\,T_\mathrm{eff}^4}
   {G\,M_\ast\,M_\mathrm{e}}
   \left( \frac{I_\mathrm{e}(n,x_0)}{I_\mathrm{m}(n,x_0)} \right)^{-1}.
\label{dr_env}
\end{eqnarray}
If the behaviour of $T_\mathrm{eff}$ is known, Eq.~(\ref{dr_env}) can be numerically
integrated to obtain the evolution of the envelope radius.


\begin{thebibliography}{}

   \bibitem[1994]{af94}
Alexander, D. R. \& Ferguson, J. W. 1994, \apj, 437, 879

   \bibitem[1976]{bs76}
Bath, G. T. \& Shaviv, G. 1976, \mnras, 175, 305

   \bibitem[2003]{bond}
Bond, H. E., Henden, A., Levay, Z. G., et~al. 2003, \nat, 422, 405

   \bibitem[1993]{bb}
Brand, J. \& Blitz, L. 1993, \aap, 275, 67

   \bibitem[2002]{brown}
Brown, N. J. 2002, IAU Circ., 7785

   \bibitem[1957]{chandra}
Chandrasekhar, S. 1957, An Introduction to the Study of Stellar Structure,
Dover Publications, Inc.

   \bibitem[2003]{lisa03}
Crause, L. A., Lawson, W. A., Kilkenny, D., et~al. 2003, \mnras, 341, 785

   \bibitem[2005]{lisa04}
Crause, L. A., Lawson, W. A., Menzies, J. W., \& Marang, F. 2005, \mnras, 358, 1352

   \bibitem[1998]{debin}
Dehnen, W. \& Binney, J. J. 1998, \mnras, 298, 387

   \bibitem[2002]{desmun}
Desidera, S. \& Munari, U. 2002, IAU Circ., 7982

   \bibitem[1985]{draine}
Draine, B. T. 1985, \apjs, 57, 587

   \bibitem[2002]{henden}
Henden, A., Munari, U. \& Schwartz, M. B. 2002, IAU Circ., 7859

   \bibitem[2002]{henmun}
Henden, A. \& Munari, U. 2002, IAU Circ., 7958

   \bibitem[1965]{iben}
Iben, I. 1965, \apj, 141, 993

   \bibitem[1966]{johnson}
Johnson, H. L. 1966, \araa, 4, 193

   \bibitem[2002]{kaufl}
Kaeufl, H. E., Lucarto, Kerber, F. \& Hijligers, B. 2002, IAU Circ., 7831
                                                                                  
   \bibitem[2002]{kimes}
Kimeswenger, S., Lederle, C., Schmeja, S. \& Armsdorfer, B. 2002,
\mnras, 336, L43

   \bibitem[2004]{kipper}
Kipper, T., Klochkova, V. G., Annuk, K., et al. 2004, \aap, 416, 1107

   \bibitem[2002]{kolev}
Kolev, D., Miko\l{}ajewski, M., Tomov, T., et~al. 2002, Collected Papers Physics,
Shumen Univ. Press, p.147

   \bibitem[1983]{koor}
Koornneef, J. 1983, \aap, 128, 84
                                                                                  
   \bibitem[2005]{lane}
Lane, B. F., Retter, A., Thompson, R. R. \& Eisner, J. A. 2005, \apj, 622, L137

   \bibitem[1970]{lee}
Lee, T. A. 1970, \apj, 162, 217
                                                                                  
   \bibitem[2003]{lynch03}
Lynch, D. K., Russell, R. W. \& Polomski, E. 2003, IAU Circ., 8221

   \bibitem[2004]{lynch04}
Lynch, D. K., Rudy, R. J., Russell, R. W., et al. 2004 \apj, 607, 460

   \bibitem[2002a]{mundes}
Munari, U., Desidera, S. \& Henden, A. 2002a, IAU Circ., 8005

   \bibitem[2002b]{munhen}
Munari, U., Henden, A., Corradi, R. M. L. \& Zwitter, T. 2002b, in Classical
Nova Explosions, eds. M. Hernanz \& J. Jose, AIP Conference Proceedings,
vol.637, p.52

   \bibitem[2002c]{munari}
Munari, U., Henden, A., Kiyota, S., et~al. 2002c, \aap, 389, L51

   \bibitem[2005]{mun05}
Munari, U., Henden, A., Vallenari, A., et al., 2005, \aap, 434, 1107

   \bibitem[2003]{osiwal}
Osiwa\l{}a, J. P., Miko\l{}ajewski, M., Tomov, T., et~al. 2003, in Symbiotic
Stars Probing Stellar Evolution, eds. R. L. M. Corradi, J. Miko\l{}ajewska
\& T. J. Mahonney (San Francisco: ASP), p.240

   \bibitem[1982]{sk}
Schmidt-Kaler, Th. 1982, Landolt-B\"ornstein: Numerical Data and Functional
Relationships in Science and Technology, eds. K. Schaifers \& H. H. Voigt
(Springer-Verlag, Berlin), VI/2b

   \bibitem[2003]{soktyl}
Soker, N. \& Tylenda, R. 2003, \apjl, 582, L105

   \bibitem[2003]{tapia}
Tapia, M. \& Persi, P. 2003, IAU Circ., 8241

   \bibitem[2000]{tokun}
Tokunaga, A. T. 2000, in Allen's
Astrophysical Quantities, the 4th edition, ed. A. N. Cox 
(Springer-Verlag, New York) p. 143

   \bibitem[2004]{tyl}
Tylenda, R. 2004, \aap, 414, 223

   \bibitem[2005a]{tcgs}
Tylenda, R., Crause, L., G\'orny, S. K. \& Schmidt, M. R. 2005a, \aap, accepted
(astro-ph/0412205)

   \bibitem[2005b]{tss}
Tylenda, R., Soker, N. \& Szczerba, R. 2005b, \aap, submitted (astro-ph/0412183)

   \bibitem[2004]{lers}
van Loon, J. Th., Evans, A., Rushton, M. T. \& Smalley. B. 2004, \aap, 427, 193

   \bibitem[2002]{wagner}
Wagner, R. M. \& Starrfield, S. G., 2002, IAU Circ., 7992

   \bibitem[2002]{watcos}
Watson, A. M. \& Costero, R. 2002, IAU Circ. 7976

   \bibitem[2003]{wisnia}
Wisniewski, J. P., Morrisson, N. D., Bjorkman, K. S. et~al. 2003, \apj, 588,
486

   \bibitem[2002]{zwimun}
Zwitter, T. \& Munari, U. 2002, IAU Circ., 7812

\end{thebibliography}
\end{document}